 \documentclass[twocolumn,a4paper]{article}
 \usepackage{etoolbox}
 \usepackage{abstract}
 \newbool{PREPRINT}
 \booltrue{PREPRINT}
\usepackage{makecell}
\usepackage[colorlinks,bookmarksopen,bookmarksnumbered,citecolor=red,urlcolor=red]{hyperref}
\usepackage[usenames]{color}
\usepackage{comment}
\usepackage{amsmath,amssymb}
\usepackage{graphicx}
\usepackage{algorithm}
\usepackage{algcompatible}
\usepackage{caption}
\usepackage{listings}
\usepackage{mathtools}
\usepackage{multicol}
\usepackage{placeins}
\usepackage{multirow}
\usepackage{subcaption}
\newcommand{\figdir}{./}

\newcommand{\order}{\mathcal{O}}
\renewcommand{\vec}[1]{\boldsymbol{#1}}

\newif\ifpreprint
\ifbool{PREPRINT}{ 
\preprinttrue
\usepackage[cm]{fullpage}
\usepackage{amsthm}
\usepackage{multicol}
\setlength\columnsep{20pt}
\usepackage[affil-it]{authblk}
\newtheorem{defn}{Definition}
\newtheorem{expl}{Example}
}{ %
\usepackage{fullpage}
\newdefinition{defn}{Definition}
\newdefinition{expl}{Example}
} 

\lstdefinelanguage[ppmd]{python}[]{python}{%
  emph={ParticleLoop,ParticleDat,PositionDat,ScalarArray,Kernel,PairLoop,Constant,State,Data,IntegratorRange}
}
\definecolor{DarkBlue}{rgb}{0.00,0.00,0.55}
\definecolor{DarkRed}{rgb}{0.55,0.00,0.00}
\definecolor{DarkGreen}{rgb}{0.00,0.55,0.00}
\definecolor{Gray}{rgb}{0.95,0.95,0.95}
\definecolor{Purple}{rgb}{0.5,0.0,0.5}
\definecolor{Bittersweet}{rgb}{1.0,0.44,0.37}
\newcommand{\DSLType}[1]{{\color{Bittersweet}#1}}
\newcommand{\DSLVar}[1]{{\color{red}#1}}
\newcommand{\PythonVar}[1]{{\color{blue}#1}}
\newcommand{\PythonConst}[1]{{\color{DarkGreen}#1}}

\newcommand{\nb}{\ensuremath{n_{\operatorname{b}}}}
\newcommand{\nnb}{\ensuremath{n_{\operatorname{nb}}}}
\newcommand{\nlcb}{\ensuremath{n_{\operatorname{lcb}}}}
\newcommand{\nuB}{\ensuremath{\nu_{\operatorname{b}}}}
\newcommand{\nuNB}{\ensuremath{\nu_{\operatorname{nb}}}}

\title{A Domain Specific Language for Performance Portable Molecular Dynamics Algorithms}

\ifbool{PREPRINT}{ 
\author[a]{William~Robert~Saunders}
\author[b]{James~Grant}
\author[a,*]{Eike~Hermann~M\"{u}ller}
\affil[ ]{University of Bath, Bath BA2 7AY, Bath, United Kingdom}
\affil[a]{Department of Mathematical Sciences}
\affil[b]{Department of Chemistry}
\affil[*]{Email: \texttt{e.mueller@bath.ac.uk}}
}{ %
\author[math]{William~Robert~Saunders}
\ead{w.r.saunders@bath.ac.uk}
\author[chem]{James~Grant}
\ead{r.j.grant@bath.ac.uk}
\author[math]{Eike~Hermann~M\"{u}ller\corref{cor1}\fnref{fn1}}
\ead{e.mueller@bath.ac.uk}
\fntext[fn1]{email: \texttt{e.mueller@bath.ac.uk}}
\cortext[cor1]{Corresponding author}
\address{University of Bath, Bath BA2 7AY, Bath, United Kingdom}
\address[math]{Department of Mathematical Sciences}
\address[chem]{Department of Chemistry}
} 

\begin{document}
\ifbool{PREPRINT}{ 
\twocolumn[
\begin{@twocolumnfalse}
\maketitle
\begin{onecolabstract}
}{%
\begin{abstract}
} 
Developers of Molecular Dynamics (MD) codes face significant challenges when adapting existing simulation packages to new hardware. In a continuously diversifying hardware landscape it becomes increasingly difficult for scientists to be experts both in their own domain (physics/chemistry/biology) and specialists in the low level parallelisation and optimisation of their codes. To address this challenge, we describe a ``Separation of Concerns'' approach for the development of parallel and optimised MD codes: the science specialist writes code at a high abstraction level in a domain specific language (DSL), which is then translated into efficient computer code by a scientific programmer. In a related context, an abstraction for the solution of partial differential equations with grid based methods has recently been implemented in the (Py)OP2 library. Inspired by this approach, we develop a Python code generation system for molecular dynamics simulations on different parallel architectures, including massively parallel distributed memory systems and GPUs. We demonstrate the efficiency of the auto-generated code by studying its performance and scalability on different hardware and compare it to other state-of-the-art simulation packages. With growing data volumes the extraction of physically meaningful information from the simulation becomes increasingly challenging and requires equally efficient implementations. A particular advantage of our approach is the easy expression of such analysis algorithms. We consider two popular methods for deducing the crystalline structure of a material from the local environment of each atom, show how they can be expressed in our abstraction and implement them in the code generation framework. 
\ifbool{PREPRINT}{ 
\end{onecolabstract}
\textbf{keywords}:
\newcommand{\sep}{, }
}{
\end{abstract}
\begin{keyword}
} 
Molecular Dynamics\sep Domain Specific Language\sep Performance Portability\sep Parallel Computing\sep GPU
\ifbool{PREPRINT}{ 
\\[1ex]
}{
\end{keyword}
\maketitle
}
\ifpreprint
\end{@twocolumnfalse}]{}
\else
\fi

\section{Introduction}
Molecular Dynamics (MD) codes such as NAMD \cite{Nelson1996,Phillips2005}, LAMMPS \cite{Plimpton1995}, GROMACS \cite{Berendsen1995,Pronk2013} and DL-POLY \cite{Smith1996,Todorov2006} are important computational tools for understanding the fundamental properties of physical, chemical and biological systems. They can be used to verify phenomenological theories about atomistic interactions, understand complex bio\-molecules \cite{Karplus1990} and self assembly processes \cite{Rapaport2004}, replace costly laboratory experiments and allow access to areas of  parameter space which are very difficult to reproduce experimentally. For example, simulations can be run at high pressures and temperatures found in stellar atmospheres \cite{Horowitz2011}, or for dangerous substances, such as radioactive materials (see e.g. \cite{Williams2015}). Classical MD codes simulate a material by following the time evolution of a large number of particles which obey the laws of classical physics (in particular Newton's laws \cite{Newton1846}) and interact via phenomenological potentials. To extract meaningful information, the state of the system (i.e. the distribution of particle positions and velocities) has to be analysed, for example by calculating pairwise distribution functions. Information on the crystalline structure of a material can be derived by inspecting the local environment of each particle \cite{Steinhardt1983,Honeycutt1987,Stukowski2012}.

In order to study systems at physically relevant length- and timescales and to produce statistically converged results, modern codes typically run in parallel on state-of-the art supercomputers \cite{Phillips2005}. With the recent rise of novel manycore chips, such as GPU and Xeon Phi processors, several popular MD simulation packages have been successfully adapted to those new architectures, see e.g. \cite{Stone2007,Anderson2008,Brown2011,Brown2012,Abraham2015,Glaser2015}. However, developers of MD codes face significant challenges: adapting and optimising existing codes requires not only a deep understanding of the physics and chemistry of the simulated system, but also detailed knowledge of the rapidly evolving hardware. To name just a few complications, GPUs have a complex memory hierarchy (host/device memory, shared memory and local registers) and any data access has to be coalesced to avoid unnecessary data movement. Write conflicts have to be avoided in threaded implementations on manycore chips and recent CPUs, such as the Intel Haswell and Broadwell chip, only run at peak performance if the code can be vectorised. Since in practice it is rare for a chemist/physicist to possess the skills for optimising code on this level, it can be very challenging to port MD software to a new architecture and maintain its performance in a rapidly evolving hardware landscape. To address this fundamental issue, we describe an approach based on the idea of a ``Separation of Concerns'' between the domain specialist and scientific programmer. By using a suitable abstraction, both the scientific capabilities and computational performance can be improved independently.

\paragraph{DSLs for grid-based PDE solvers}
Very similar issues have been faced by developers of grid-based solvers for partial differential equations (PDEs). The key observation there was that the fundamental and computationally most expensive operations can be expressed in terms of a suitable abstraction: the algorithms (e.g. explicit time stepping methods or iterative solvers for elliptic PDEs) can be formulated as the repeated iterations over a set of grid entities (cells, vertices, faces, edges), each of which can hold information, such as a local field value. This expression of the algorithm in a Domain Specific Language (DSL) simplifies the implementation significantly: once the domain-specialist has expressed the code in terms of those basic operations at the correct abstraction level and encapsulated any data in the corresponding fundamental data structures, a computational scientist can implement and optimise the code on a particular architecture. 

By introducing the correct abstraction, only a small set of typical loops, which can be pa\-ra\-me\-tri\-sed over the set of input and output data, has to be considered. This concept has been applied very successfully in the development of the performance-portable OP2 library \cite{Bertolli2012,Giles2013}, which allows the execution of finite element and finite volume codes on a range of architectures. As de\-mon\-stra\-ted in \cite{Giles2011,Bertolli2012,Giles2013,Reguly2016}, the code achieves excellent performance on CPUs, GPUs and Xeon Phi processors. Similar techniques for structured grids have been used to develop the C++ based STELLA grid library for the COSMO numerical weather forecast model \cite{Gysi2014}. DSLs for highly efficient stencil computations on GPUs have also been described in \cite{Maruyama2011,Hu2016}.

Recently OP2 was re-implemented in Python as the PyOP2 \cite{Rathgeber2012} framework. In PyOP2 the science user specifies the computationally most expensive operations as a set of small kernels written in C. Using code generation techniques, those kernels are then compiled and executed on a particular architecture. By employing just-in-time compilation, the kernels are launched from a high-level Python code which implements the overall solver algorithm. The performance of the resulting code is on a par with that of monolithic Fortran- or C- implementations.
\paragraph{A new DSL for MD simulations}
In this paper we describe a similar DSL approach for molecular dynamics simulations. The fundamental operation we consider is a two-particle kernel: the user implements a short C-code which is executed for each combination of particle pairs in the simulation. This kernel can modify any properties stored on those particles. A classic example is the force calculation: for each pair of particles, the force (output) is calculated as a function of the two particle positions (input). This local operation can be expressed in a few lines of C-code. The code is then executed over all particle pairs, using the optimal algorithm for a particular hardware and the nature and size of the problem. For example, on a CPU architecture, cell-list or neighbour-list methods can be used, whereas on GPU a neighbour-matrix approach as in \cite{Rapaport2011} might be more suitable. 
Those details of the kernel execution, however, are of no interest for the science developer who can focus on (i) the implementation of the local kernel and (ii) the overall algorithm which orchestrates the kernel calls in an outer timestepping loop.

To achieve this we developed a Python-based code generation system which creates and compiles fast, architecture dependent wrapper code to execute the C-kernel over all particle pairs. Our approach is shown schematically in Fig. \ref{fig:framework_structure}.
\begin{figure}
  \begin{center}
    \includegraphics[width=\linewidth]{\figdir/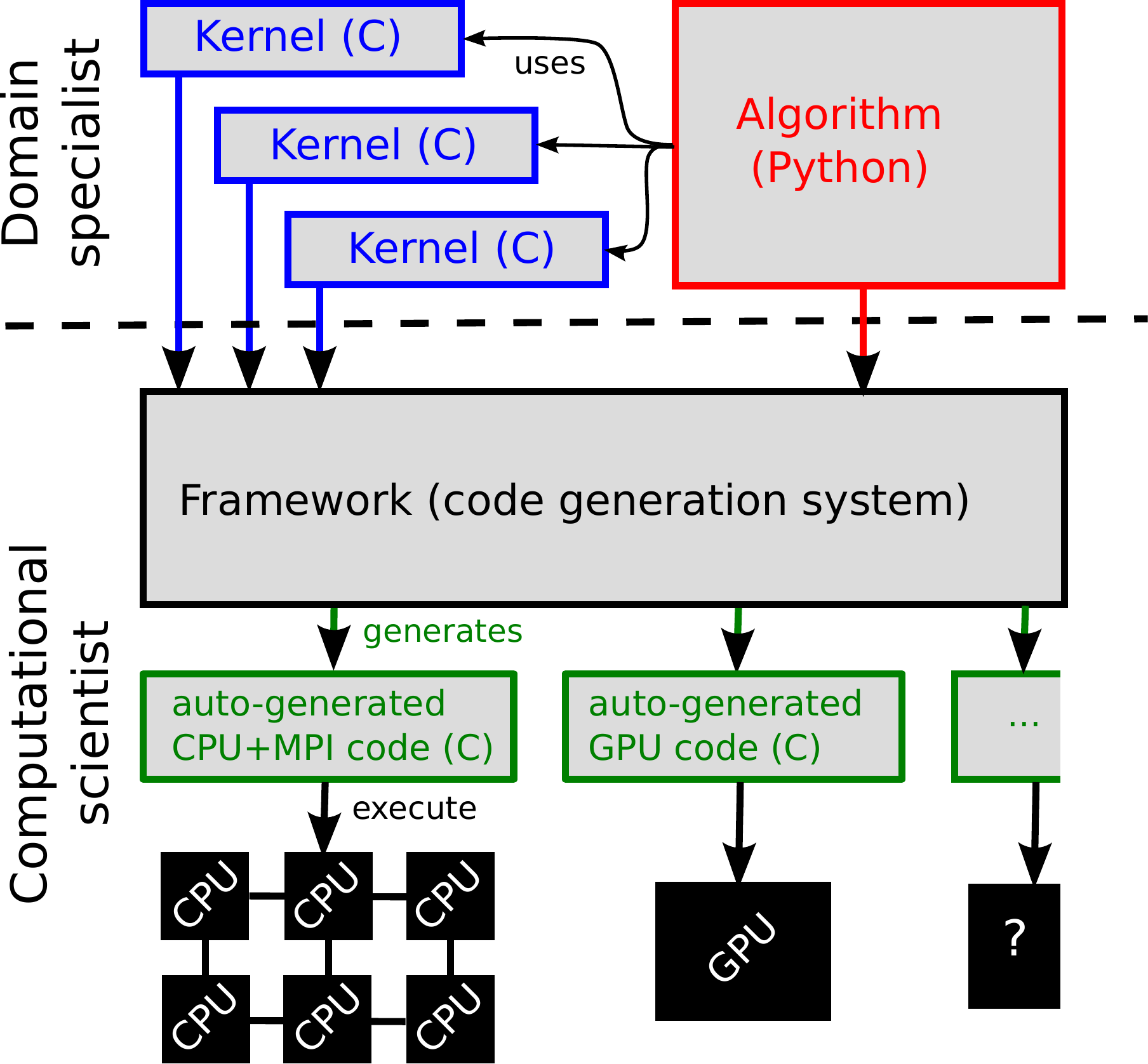}
    \caption{Structure of the code generation framework. The ``Separation of concerns'' between the domain specialist user and computational scientist is indicated by the dashed horizontal line.}
    \label{fig:framework_structure}
  \end{center}
\end{figure}
By using Python as a high-level language, looping algorithms such as the Velocity Verlet method \cite{Verlet1967} (see also e.g.\ \cite{Allen1989,Frenkel2001}) for timestepping or advanced thermostats \cite{Jia2007,Leimkuhler2011} can be implemented very easily, while still generating fast code for the computationally expensive particle loops.

In the following we describe a proof-of-concept implementation of the DSL and concentrate on short-range two-particle kernels, i.e.\ kernels which are only executed for particles which are separated by no more than a specified cutoff distance. We demonstrate that for a Lennard-Jones benchmark we achieve performance similar to state-of-the-art simulation tools such as DL-POLY and LAMMPS.

While many atomistic models require the calculation of long range forces and intra-molecular interactions, systems containing only short range interactions remain actively studied, particularly in problems in soft matter and nucleation see e.g. \cite{Radu2017,Razali2017}. In a separate paper \cite{Saunders2017a} we report on the implementation of a particle-Ewald method \cite{Ewald1921} for electrostatic forces in our framework. As discussed in Section \ref{sec:Conclusion}, more advanced long range algorithms and further generalisations of the framework to support multiple species and bonded interactions for molecules will be implemented in the future.

We stress, however, that our approach is not limited to force calculations. To extract meaningful information from a simulation, the results have to be analysed. With growing problem sizes and data volumes, this step becomes computationally expensive and requires efficient and parallel implementations. Below we consider two methods for analysing local environments which can be used to classify the crystalline phase of a material: the bond order analysis in \cite{Steinhardt1983} and common neighbour analysis in \cite{Honeycutt1987} (see also \cite{Stukowski2012} for an overview of other analysis methods). In the traditional approach, the user would run the simulation with an existing MD package and then write post-processing code to extract physically meaningful information from the output. However, in contrast to the MD code itself, parallelising this analysis code or porting it to a different architecture is often too time consuming to be feasible. As we will demonstrate below, the fundamental kernels for various common analysis methods can be expressed in our framework. This implies that optimised and parallel code is automatically generated for this important stage of the simulation workflow.

A high-level approach for introducing new algorithms to existing MD packages has been realised in the PLUMED \cite{Bonomi2009} and MIST \cite{Bethune2016} libraries. They are written as plug-ins to well-established codes and introduce free energy methods and alternative integrators respectively. However, this approach still requires the underlying MD code to be implemented efficiently in the first instance. Similar high-level Python interfaces are provided by OpenMM \cite{Eastman2013} and HOOMD-blue \cite{Anderson2008}; in those two cases the underlying code is part of the package itself. Using these interfaces both OpenMM and HOOMD-blue allow the user to control a simulation and access available particle data through calls to the underlying library. A Python based DSL for MD simulations is described in \cite{Cickovski2007}: the Molecular Dynamics Language (MDL) provides data structures for particle vectors and allows the easy construction of new integrators via Python classes. It also provides an interface to existing algorithms from the ProtoMol packages and support for reading MD configuration file formats. The main purpose of MDL is to provide a scripting environment for rapid prototyping of new timestepping algorithms. Although there is support for MPI parallelism, the main focus is not on performance or portability. While using optimised C++ implementations from ProtoMol, in contrast to our approach there is no code generation.

Many MD libraries support the implementation of custom interactions by either providing a mechanism that interpolates tabulated values to produce a potential, or a plugin system that allows users to write and compile extensions that implement the desired interaction. The OpenMM Python interface allows a custom potential to be described in symbolic form. Based in this, OpenMM will automatically generate GPU code by using symbolic differentiation and code generation. The resulting code is compiled at runtime through the OpenCL compiler.

However in all cases (with the exception of kernel code generation in OpenMM) the primary aim of the provided Python interface is to simplify access to functionality in an underlying C++ or Fortran code, i.e. Python acts as a ``glue'' for combining existing functionality. If a desired simulation or technique cannot be described within the Python interface for the library, the user needs to program extensions for the specific MD package. In contrast, our approach is more invasive and allows the expression of both the high-level algorithm and low level kernel in one code. We support general kernels, which are not restricted to force calculations that can be expressed in mathematical form.
\paragraph{Structure} This paper is organised as follows: in Section \ref{sec:Abstraction} we introduce the fundamental abstractions and data structures used in our approach. The implementation of the abstractions in a Python library and code generation techniques for different architectures are discussed in Section \ref{sec:Implementation}. In Section \ref{sec:analysis} we show how fairly complex structure analysis techniques based on bond order- and common neighbour- analysis can be expressed in our abstraction and explain how they can be added to the simulation. To demonstrate the performance of the generated code, we compare runtime and scalability to other popular MD packages both on MPI-parallel clusters and for GPUs in Section \ref{sec:Results}. Here we also show output of the structure analysis algorithms described in Section \ref{sec:analysis}.
We conclude and outline ideas for further developments in Section \ref{sec:Conclusion}.
\section{Abstraction}\label{sec:Abstraction}
We begin by formulating the key operations which are required to develop a generic MD code. If the domain specialist (computational physicist or chemist) can express their algorithms in terms of those operations, then the code can be implemented in a performance portable way in the high-level Python framework described in Section \ref{sec:Implementation}.

Throughout this paper we assume that we want to simulate and analyse a collection of $N\gg 1$ particles. Let each particle with global index $i\in\{0,1,2,\dotsc,N-1\}\equiv \mathcal{N}$ have a set of properties $\pi$ such that $\pi^{(i)}_{r}$ is the value of the $r$-th property on particle $i$. Each particle has exactly $M$ properties, i.e. $r\in [0,M-1]\equiv\mathcal{M}$. Properties can, for example, be the particle's position and momentum vector, its charge or the particle index. In addition there can be $M^g$ \textit{global} properties $\pi^g_{r^g}$ with $r^g\in [0,M^g-1]\equiv\mathcal{M}^g$. Typical global properties might be the total kinetic energy or the radial distribution function (represented as a vector $\vec{R}$ with entries $R_i$ which count the average number of particles in each distance interval $[r_i,r_{i+1}]$).

Operations which involve one or more particles are described in the following three definitions:
\begin{defn}
  A \textit{Particle Loop} is an operation which for each particle $i\in \mathcal{N}$ reads properties $\pi^{(i)}_{r}$ with $r\in\mathcal{M}_R\subset \mathcal{M}$ and writes properties $\pi^{(i)}_{s}$ with $s\in\mathcal{M}_W\subset \mathcal{M}$. The operation can also read global properties $\pi^g_{r^g}$ with $r^g\in\mathcal{M}^g_R\subset\mathcal{M}^g$ and write $\pi^g_{s^g}$ with $s^g\in\mathcal{M}^g_W\subset\mathcal{M}^g$ such that the final value of these global properties is independent of the order in which it loops over the particles.
\end{defn}
\begin{expl}
\textit{Kinetic energy calculation.} To calculate the total kinetic energy, we loop over all particles $i$ and add $\frac{1}{2} m^{(i)} \sum_{k=0}^{d-1}(v^{(i)}_{k})^2$ to the global variable $K$. The particle properties considered in this example are the mass $m^{(i)}$ and the three components $v^{(i)}_{k}$, $k=0,1,2$ of the particle's velocity vector $\vec{v}^{(i)}$.
\end{expl}
\begin{defn}
  A \textit{Particle Pair Loop} is an operation which for all particle pairs $(i,j)\in \mathcal{N}\times\mathcal{N}$ reads properties $\pi^{(i)}_{r}$ and $\pi^{(j)}_{r}$ with $r\in \mathcal{M}_R\subset \mathcal{M}$ and modifies properties $\pi^{(i)}_{s}$ with $s\in\mathcal{M}_W\subset\mathcal{M}$ such that the result is independent of the order of execution. The kernel can also read global properties $\pi^g_{r^g}$ with $r^g\in\mathcal{M}^g_R\subset\mathcal{M}^g$ and write $\pi^g_{s^g}$ with $s^g\in\mathcal{M}^g_W\subset\mathcal{M}^g$ such that the result does not depend on the order in which the loop is executed over all particle pairs.
\end{defn}
\begin{expl}
\textit{Force Calculation.} The most obvious example of a Particle Pair Loop is the force calculation. Here each particle has six relevant properties, namely the three entries of its position vector and the three entries of the force exerted on the particle by all other particles. For each particle pair the total force on the first particle is incremented by the interaction force $\vec{f}(\vec{r}^{(i)},\vec{r}^{(j)})$ which depends on the relative position of the particles, i.e.\ the three position properties $r^{(i)}_k$ for $k=0,1,2$ are read and the three force properties $F^{(i)}_k$ are incremented as $F^{(i)}_k \mapsto F^{(i)}_k + f_k(\vec{r}^{(i)},\vec{r}^{(j})$.
\end{expl}
\begin{defn}
  A \textit{Local Particle Pair Loop} is a Particle Pair Loop which is only executed for particles which are separated by no more than a specified cutoff distance $r_c$.
\end{defn}
\begin{expl}
\textit{Local environment.} Suppose that each atom can be in one of two possible states. For every atom we want to count the number of other atoms in the same state which are up to a distance $r_c$ away. In this case each particle would have five properties, namely the three entries of the position vector, the state of the atom and the number of atoms in the same state in the local environment. For each pair of atoms the Particle Pair Kernel would first check whether they are less than $r_c$ apart by calculating the distance $|\vec{r}^{(i)}-\vec{r}^{(j)}|$ between the particle positions. If this is the case, and both particles are in the same state, the counter for the number of same-state atoms is increased.
\end{expl}
Further examples will be given in Section \ref{sec:analysis} where we show how the bond order analysis in \cite{Steinhardt1983} and a common neighbour analysis \cite{Honeycutt1987} can be expressed as Particle- and Particle Pair- Loops. The Particle Pair Loop can be easily generalised to a loop involving $k>2$ particles for multiparticle forces.

Note that the computational complexity of a \textit{Local Particle Pair} loop is $\order(N\cdot N_{\text{local}})$ where $N_{\text{local}}=(4/3)\pi r_c^3 \rho$ is the average number of local neighbours. Since, for constant density $\rho$, the number $N_{\text{local}}$ is constant and relatively small, the computational complexity is $\order(N)$ and therefore significantly smaller than the $\order(N^2)$ complexity of a \textit{Particle Loop}.
\paragraph{Comment on Newton's third law} For most physically relevant interactions the force on the first particle of the pair is equal and opposite to the force acting on the second particle. Hence, instead of looping over all $N(N-1)$ unordered pairs $(i,j)$, one could also only loop over the $N(N-1)/2$ ordered pairs with $i<j$, calculate the force once and update it on both particles. Naively this should lead to a speedup of a factor of two. However, it introduces write conflicts in a (shared memory) parallel implementation. While those can be avoided by adding suitable atomic statements or using a colouring approach, the more serious issue is that it prevents automatic vectorisation. When writing back to memory, the compiler has to assume that there could be aliasing between particle data (from the compiler's point of view two of the neighbours of each particle could be identical), and will not generate vectorised code. This can be overcome by suitable clustering of the neighbour lists \cite{Pall2013} or blocking of the pair lists \cite{Mangiardi2017} and explicit vector load/store operations. Note, however, that the authors of \cite{Pall2013} use architecture dependent vector instructions in their kernels, which we want to avoid to achieve portability.

Here we do not use any of those approaches and rely on automatic vectorisation, which works well if we only write to the first particle in each pair. In summary we observe that the factor of two which could be gained by using Newton's third law is more than offset by the advantages of vectorisation and we find that the code is faster overall if we loop over all ordered pairs and only write to the first particle. As will be demonstrated in Section \ref{sec:comparison}, for short range forces we achieve equal or better performance than other common MD packages. If necessary, it would of course be possible to implement a version of the pair looping mechanism which exploits Newton's second law in our code generation framework and improvements such as those described in \cite{Pall2013,Mangiardi2017} could be considered in future extensions.
\section{Implementation}\label{sec:Implementation}
The operations identified in the previous section are the computationally most expensive components of an MD simulation. We now describe their efficient parallel implementation in a code generation framework. From the discussion above it should be clear that our framework will have to provide (1) data structures to represent particle properties $\pi^{(i)}_{r}$ as well as global properties $\pi^g_r$ and (2) mechanisms for executing Particle- and Particle Pair-Loops. The following choices are inspired by the PyOP2 \cite{Rathgeber2012} data structures and execution model. An implementation of the framework described in this section can be found at:\begin{center}\url{https://bitbucket.org/wrs20/ppmd}\end{center}All results in this paper were obtained with the release available as \cite{ppmd_release}.
\subsection{Data structures}
Particle properties $\pi^{(i)}_{r}$ are represented as instances of a \verb!ParticleDat! class. This class is a wrapper around a two-dimensional \verb!numpy! array, where the first index labels the particle $i$ and the second corresponds to the property index $r$. Similarly we provide storage for global data shared by all particles in a \verb!ScalarArray! class.

For convenience and to support different data types, we do not collect all properties into a single \verb!ParticleDat! (or \verb!ScalarArray!), but rather allow several \verb!ParticleDat!s and \verb!ScalarArray!s instances which can be named by the user. For example, consider a simulation with particles which have three dimensional position and momentum vectors $\vec{r}^{(i)},\vec{v}^{(i)}\in\mathbb{R}^3$ and a species index $S^{(i)}\in \mathbb{N}$. We also store the total kinetic- and potential energies $KE, PE\in \mathbb{R}$. This set of local and global properties would be implemented as shown in Listing \ref{lst:particle-dats}.
\begin{figure}
\begin{minipage}{\linewidth}
\begin{lstlisting}[language={[ppmd]{python}}, label=lst:particle-dats,caption={Data structure initialisation}]
x = ParticleDat(ncomp=3,dtype=c_double)
v = ParticleDat(ncomp=3,dtype=c_double)
S = ParticleDat(ncomp=1,dtype=c_int,
                initial_value=0)
KE = ScalarArray(ncomp=1,
                 dtype=c_double,
                 initial_value=0.0)
PE = ScalarArray(ncomp=1,
                 dtype=c_double,
                 initial_value=0.0)
\end{lstlisting}
\end{minipage}
\end{figure}

The underlying \verb!numpy! array can be accessed as the \verb!ParticleDat.data! property; however the ``getitem'' and ``setitem'' methods have been overloaded to automatically mark the \verb!ParticleDat! as ``dirty'' if the internal data has been modified directly by the user. This is important in parallel implementations based on a domain decomposition approach, where data owned by neighbouring processors is duplicated in a ``halo'' region. If ``dirty'' data is used subsequently in a loop, a exchange of halo data will be triggered automatically and ensures that data is consistent between processors. The interface to the stored data is identical for both CPU- and GPU- \verb!ParticleDat! data structures. When accessing data stored in a \verb!ParticleDat! stored on the GPU in device memory, ``getitem'' and ``setitem'' calls will automatically trigger data copies between host- and device-memory. The correct architecture is chosen at the beginning of the Python script by setting aliases for the appropriate objects as shown in Listing \ref{lst:GPU-switch}.
\begin{figure}
\begin{minipage}{\linewidth}
\begin{lstlisting}[language={[ppmd]{python}}, label=lst:GPU-switch,caption={Switching between CPU and GPU implementation}]
import ppmd as md
# Set USE_CUDA to True or False
if not USE_CUDA:
    Data = md.data
    State = md.state.State
    ParticleLoop =
      md.loop.ParticleLoop
    PairLoop =
      md.pairloop.PairLoopNeighbourListNS
else:
    Data = md.cuda.cuda_data
    State = md.cuda.cuda_state.State
    ParticleLoop =
      md.cuda.cuda_loop.ParticleLoop
    PairLoop =
      md.cuda.cuda_pairloop. \
        PairLoopNeighbourListNS

PositionDat = Data.PositionDat
ParticleDat = Data.ParticleDat
ScalarArray = Data.ScalarArray
\end{lstlisting}
\end{minipage}
\end{figure}
\subsection{Particle Pair Loops}\label{sec:ParticlePairLoops}
In addition to data structures, an execution model is required to launch the computational kernel over all particle pairs. For this, the user writes a brief C-kernel which describes how the properties of the two particles involved in the interaction are modified. In addition, the \verb!ParticleDat!s which are operated on have to be passed explicitly to the pair looping mechanism. For each \verb!ParticleDat! an access descriptor describes whether the property is read from or written to. The allowed access descriptors are \verb!READ! (property is only read), \verb!WRITE! (property is only written to), \verb!RW! (property is read and written), \verb!INC! (property is incremented) and \verb!INC_ZERO! (identical to \verb!INC! except the values are set to zero before the kernel is launched); see also Tab. \ref{tab:DSL_access} for a summary. Since the code generation system does not inspect the C-kernel provided by the user, this information allows the looping system to handle read- and write- access to particle properties in a parallel setting. For example, in a distributed memory implementation, before the execution of the loop halo regions have to be updated for all variables which have a \verb!READ! access descriptor. Similarly, if a particle has \verb!WRITE! or \verb!INC! access, in a threaded implementation write conflicts have to be avoided by generating atomic write statements or employing suitable colouring (see for example the layer algorithm described in \cite{Rapaport1988}). In addition to \verb!ParticleDat!s, global variables (represented as \verb!ScalarArray!s) can be passed to the kernel with the same access descriptors. To treat numerical constants which do not change during the kernel execution, each kernel can also be be passed a list of \verb!Constant! objects. Any instances of \verb!Constant! variables in a kernel are replaced by their numerical values at compile time; this allows the compiler to make additional optimisations, for example by exploiting static loop bounds.

As an (fictitious) example, imagine that on each particle we store the properties $a$ (which has $d=3$ components) and $b$ (which has one component).
For all particles $i$ we carry out the operation which calculates
\begin{equation}
  b^{(i)} = \sum_{\text{all pairs $(i,j)$}} \sum_{r=0}^{d-1} \left(a^{(i)}_{r}-a^{(j)}_{r}\right)^2 \label{eqn:simple_op}
\end{equation}
and updates the global sum
\begin{equation}
  S^{g} = \sum_{\text{all pairs $(i,j)$}} \sum_{r=0}^{d-1} \left(a^{(i)}_{r}-a^{(j)}_{r}\right)^4. \label{eqn:simple_op_global}
\end{equation}
A Particle Pair loop which performs this operation can be implemented as shown in Listing \ref{lst:simple-kernel}. The execution over all particle pairs is illustrated schematically in Fig. \ref{fig:pairwise_schematic}.\\
\begin{figure}
\begin{minipage}{\linewidth}
\begin{lstlisting}[language={[ppmd]{python}}, label=lst:simple-kernel,caption={Python code for executing the operations in Eqs. (\ref{eqn:simple_op}) and (\ref{eqn:simple_op_global}) over all particle pairs.}]
# dimension
dimension=3

# number of particles
npart=1000

# Define Particle Dats
a = ParticleDat(npart=npart,
                ncomp=dimension,
                dtype=c_double)
b = ParticleDat(ncomp=1,
                npart=npart,
                initial_value=0.0,
                dtype=c_double)
S = ScalarArray(ncomp=1,
                initial_value=0.0,
                dtype=c_double)

kernel_code='''
  double da_sq = 0.0;
  for (int r=0;r<dimension;++r) {
    double da = a.i[r]-a.j[r];
    da_sq += da*da;
  }
  b.i[0] += da_sq;
  S += da_sq*da_sq;
'''

# Define constants passed to kernel
kernel_consts = (Constant('dimension',
                          dimension),)

# Define kernel
kernel = Kernel('update_b',
                kernel_code,
                kernel_consts)
  
# Define and execute pair loop
pair_loop = PairLoop(kernel=kernel,
                     {'a':a(access.READ),
                      'b':b(access.INC),
                      'S':S(access.INC)})
pair_loop.execute()
\end{lstlisting}
\end{minipage}
\end{figure}
\begin{figure}
  \begin{center}
    \includegraphics[width=1.0\linewidth]{\figdir/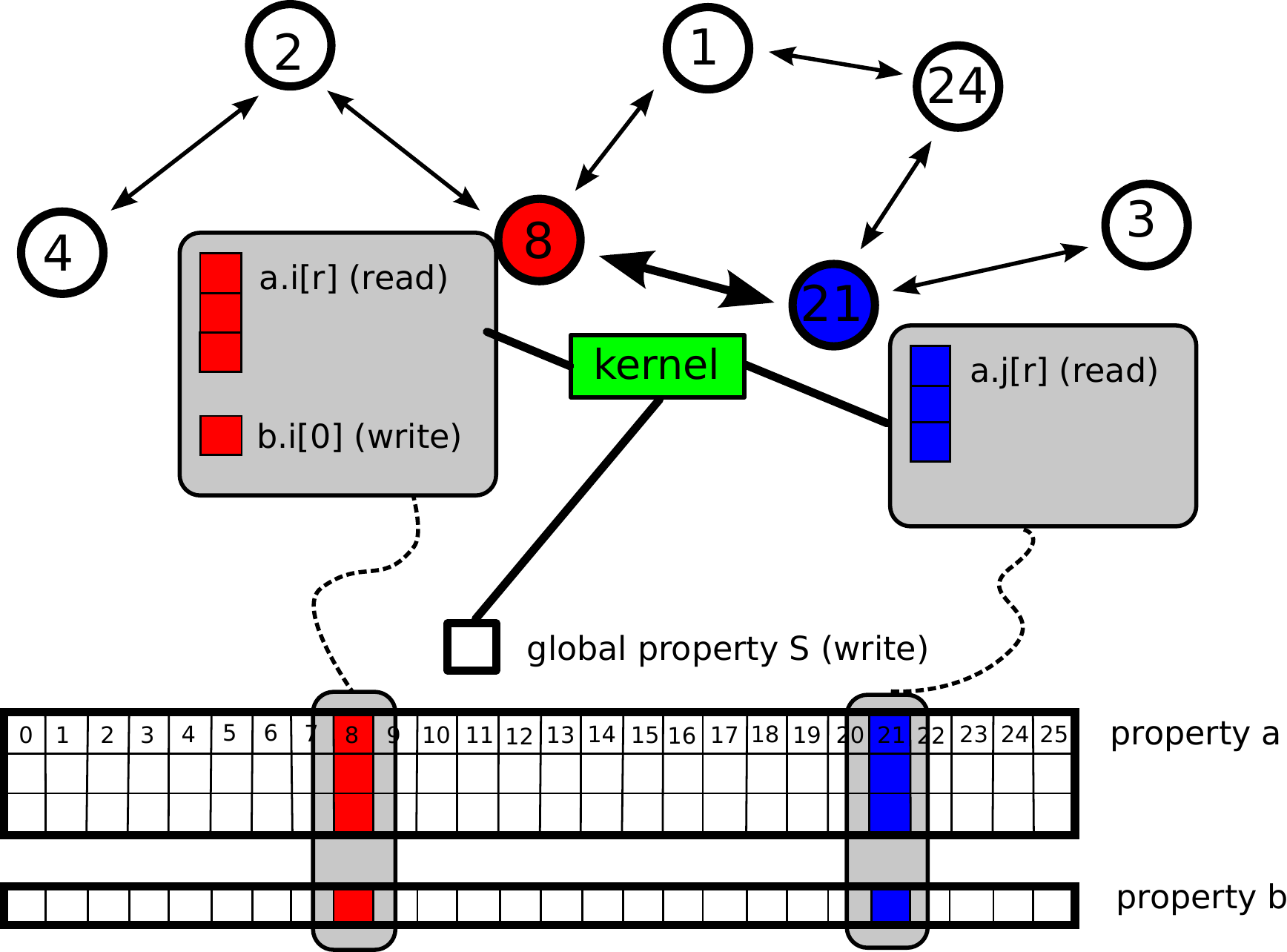}
    \caption{Pairwise kernel for executing the operation in Eqs. (\ref{eqn:simple_op}) and (\ref{eqn:simple_op_global}) over all particle pairs.}
    \label{fig:pairwise_schematic}
  \end{center}
\end{figure}
Inside the Particle Pair Loop the two involved particles are accessed as the \verb!.i! and \verb!.j! component of a structure, and the names of the \verb!ParticleDat!s are given in the dictionary which is passed as the second argument to the \verb!PairLoop! constructor. For example, the $r$-th component of the first particle is accessed as \verb!a.i[r]!. This C-variable automatically points to the correct position in the \verb!numpy! array which holds the \verb!ParticleDat! values. Particle Loops are conceptually very similar and can be implemented in the same way. While the simple example above aims to illustrate the key concepts of our approach, we also describe the implementation of a complete Lennard-Jones benchmark with Velocity-Verlet integrator in Section \ref{sec:Results}. The C- and Python-code for executing the force calculation in this case is given in Listings \ref{lst:LJ-kernel} and \ref{lst:LJ-loop} in \ref{sec:ForceCalculation}.

We note that the code in Listing \ref{lst:simple-kernel} resembles what would be written in PyOP2 to implement a loop over a set of mesh entities. In PyOP2 the fundamental data types are called \verb!Dat! and \verb!GlobalDat!. A \verb!Dat! object represents data which is associated with topological entities of the mesh, for example the average value of a field in each grid cell. A \verb!GlobalDat! variable contains globally available data. The main difference is that PyOP2 loops over a particular static set of topological entities and can access data on other related entities which are specified via indirection maps. Those indirection maps are provided as additional arguments to the \verb!Dat! dictionary of the looping class. An important difference is that the indirection maps in PyOP2 have a fixed ``arity'', i.e. each unknown depends on a fixed number of other unknowns. In contrast, in an MD code, the number and identity of nearest neighbours of each particle varies throughout the simulation.
In a parallel MD code the distribution of particles over processors also changes over time, and this requires additional parallel communication.
\subsection{Domain Specific Language}
The key Python classes for representing MD specific data objects in our embedded DSL are summarised in Tab. \ref{tab:DSL_data}. The looping classes which are used to modify those fundamental objects according to the \textit{Particle Loop} and \textit{Particle Pair Loop} operations defined mathematically in Section \ref{sec:Abstraction} are given in Tab. \ref{tab:DSL_looping}. Valid access descriptors are listed in Tab. \ref{tab:DSL_access}. For clarity instances of fundamental Python types are coloured in \PythonVar{blue}, the DSL specific classes are shown in \DSLType{orange} and instances of those classes in \DSLVar{red}. The semantics of the language have been explained in the preceeding sections.
The code strings used in the \texttt{Kernel} objects have to be legal C-code, and the particle properties can be accessed as described in Section \ref{sec:ParticlePairLoops}.

While the spectrum between pure DSLs (such as the Unified Form Language \cite{Alnaes2014}) and APIs (such as, for example, the BLAS/LAPACK libraries \cite{Lawson1979,Anderson1999}) is somewhat fluid, we argue that our approach does represent an (embedded) DSL since:
\begin{enumerate}
  \item It allows the expression of domain-specific mathematical operations (\textit{Particle}- and \textit{Particle Pair loops}) for the fundamental data objects (= particle properties).
  \item It is relatively complete in the sense that it allows the expression of key operations in MD codes; it is not restricted to the composition of high-level operations such as calls to pre-defined force terms.
  \item The user has full low-level control in the sense that they can directly manipulate the fundamental data objects in the C-kernel; this allows the implementation of complex force calculations or analysis algorithms. 
\end{enumerate}
In this sense it differs from other, more scripting-like approaches such as the PLUMED \cite{Bonomi2009} or MIST \cite{Bethune2016} libraries which mainly provide high-level APIs to existing MD packages.
\begin{table*}
\begin{center}
\begin{tabular}{|l|l|}
\hline
Description & Python Class\\
\hline
\hline
\begin{minipage}{0.5\linewidth}
Collection of properties for all particles with $d$ components per particle. All values are initialised to $x_0$ when the object is created.
\end{minipage} & 
\begin{minipage}{0.5\linewidth}
\begin{tabbing}
\texttt{\DSLType{ParticleDat}(}\=\texttt{ncomp=}\PythonVar{$d$}\texttt{,}\\
\>\texttt{dtype=\PythonConst{c\_double/c\_int/c\_long/...},}\\
\>\texttt{initial\_value=}\PythonVar{$x_0$}\texttt{)}
\end{tabbing}
\end{minipage}
\\\hline
\begin{minipage}{0.5\linewidth}
Specialisation of \DSLType{\texttt{ParticleDat}} for particle positions (see Section \ref{sec:cell_based_methods}).
\end{minipage} & 
\begin{minipage}{0.5\linewidth}
\begin{tabbing}
\texttt{\DSLType{PositionDat}(}\=\texttt{ncomp=}\PythonVar{$d$}\texttt{,}\\
\>\texttt{dtype=\PythonConst{c\_double/c\_int/c\_long/...},}\\
\>\texttt{initial\_value=}\PythonVar{$x_0$}\texttt{)}
\end{tabbing}
\end{minipage}
\\\hline
\begin{minipage}{0.5\linewidth}
Global property (not specific to individual particles) with $d'$ components; values are initialised to $y_0$.
\end{minipage} & 
\begin{minipage}{0.5\linewidth}
\begin{tabbing}
\texttt{\DSLType{ScalarArray}(}\=\texttt{ncomp=}\PythonVar{$d'$}\texttt{,}\\
\>\texttt{dtype=\PythonConst{c\_double/c\_int/c\_long/...},}\\
\>\texttt{initial\_value=}\PythonVar{$y_0$}\texttt{)}
\end{tabbing}
\end{minipage}
\\\hline
\begin{minipage}{0.5\linewidth}
Numerical constant which is replaced by its specific value in kernel, i.e. the string $L$ is replaced by the numerical value $x$ in the generated C-code. 
\end{minipage} & 
\begin{minipage}{0.5\linewidth}
\begin{tabbing}
\texttt{{\color{Bittersweet}Constant}(}\=\texttt{label=}\PythonVar{$L$}\texttt{,}\\
\>\texttt{value=}\PythonVar{$x$}\texttt{)}
\end{tabbing}
\end{minipage}
\\\hline
\begin{minipage}{0.5\linewidth}
Kernel object which can be used in one of the looping classes defined in Tab. \ref{tab:DSL_looping}. The C-source code is given as a string $S$ and any numerical constants $C_1$, $C_2$, \dots can be passed in as a list of \DSLType{\texttt{Constant}} objects.
\end{minipage} & 
\begin{minipage}{0.5\linewidth}
\begin{tabbing}
\texttt{\DSLType{Kernel}(}\=\texttt{label=}\PythonVar{$L$}\texttt{,}\\
\>\texttt{code=}\PythonVar{$S$}\texttt{,}\\
\>\texttt{constants=\DSLVar{(}}\DSLVar{$C_1,C_2,\dots,$}\texttt{\DSLVar{)} )}
\end{tabbing}
\end{minipage}
\\
\hline
\end{tabular}
\caption{Fundamental data classes of the DSL}\label{tab:DSL_data}
\end{center}
\end{table*}
\begin{table*}
\begin{center}
\begin{tabular}{|l|l|}
\hline
Description & Python Class\\
\hline
\hline
\begin{minipage}{0.5\linewidth}
Execute \DSLType{\texttt{Kernel}} object $k$ for all particles and modify particle data (\DSLType{\texttt{ParticleDat}}, \DSLType{\texttt{PositionDat}} or \DSLType{\texttt{ScalarArray}} objects) $d_1$, $d_2$, \dots. Each particle data object $d_i$ can be accessed via the corresponding label $L_i$ and has access descriptor $A_i$ defined in Tab. \ref{tab:DSL_access}.
\end{minipage} & 
\begin{minipage}{0.5\linewidth}
\begin{tabbing}
\texttt{\DSLType{ParticleLoop}(}\=\texttt{kernel=}\DSLVar{$k$}\texttt{,}\\
\>\texttt{part\_dats=\{}\=\PythonVar{$L_1$}\texttt{:}\DSLVar{$d_1$}(\DSLVar{$A_1$})\texttt{,}\\
\> \>\PythonVar{$L_2$}\texttt{:}\DSLVar{$d_2$}(\DSLVar{$A_2$})\texttt{,}\\
\>\>\dots\texttt{\} )}
\end{tabbing}
\end{minipage}\\
\hline
\begin{minipage}{0.5\linewidth}
Same as \DSLType{\texttt{ParticleLoop}}, but execute the kernel over all \textit{pairs} of particles.
\end{minipage} & 
\begin{minipage}{0.5\linewidth}
\begin{tabbing}
\texttt{\DSLType{PairLoop}(}\=\texttt{kernel=}\DSLVar{$k$}\texttt{,}\\
\>\texttt{part\_dats=\{}\=\PythonVar{$L_1$}\texttt{:}\DSLVar{$d_1$}(\DSLVar{$A_1$})\texttt{,}\\
\> \>\PythonVar{$L_2$}\texttt{:}\DSLVar{$d_2$}(\DSLVar{$A_2$})\texttt{,}\\
\>\>\dots\texttt{\} )}
\end{tabbing}
\end{minipage}\\
\hline
\end{tabular}
\caption{Fundamental looping classes of the DSL}\label{tab:DSL_looping}
\end{center}
\end{table*}
\begin{table}
\begin{center}
\begin{tabular}{|l|l|}
\hline
Description & Access Descriptor\\
\hline\hline
Read-only access & \texttt{access.READ}\\\hline
Write-only access & \texttt{access.WRITE}\\\hline
Read and write access & \texttt{access.RW}\\\hline
Incremental access & \texttt{access.INC}\\\hline
Incremental access, & \texttt{access.INC\_ZERO}\\
initialise to zero & \\
\hline
\end{tabular}
\caption{Supported access descriptors}
\label{tab:DSL_access}
\end{center}
\end{table}
\subsection{Code generation for performance-portability}

To execute a pairloop we use a code generation approach. Given the kernel and information on how data is accessed, appropriate wrapper C-code for launching the kernel over all particle pairs is generated for a particular hardware backend. This means that to target different architectures, the user has to write the kernel code only once: it is up to the code generation system (developed by a computational scientist) to execute this on a specific architecture. The implementation generates C code by first inserting the user written kernel into a pre-made template for the specified looping type, then for each passed \verb+ParticleDat+ or \verb+ScalarArray+ C code is added that matches the specified access descriptor. The result of the code generation stage is a C function which is subsequently compiled into a shared library using the C compiler defined by the user. The shared library is then loaded by the framework using the \verb+ctypes+ Python module such that it may be called directly from the Python code.

Note that the user never has to explicitly add calls to MPI routines or guarantee the correctness of the results on a threaded architecture by protecting write statements with ``atomic'' or ``critical'' keywords.

On a particular architecture different pair looping mechanisms (described below) lead to the same scientific result but can have different computational performance. Our method allows the straightforward comparison between different looping mechanism without the user intervention to modify code, a feature that could potentially be exploited to optimise performance on a problem-by-problem basis.  Since the system is aware of data dependencies between different kernel, loop fusion to reduce the amount of data movement could be implemented to further improve performance.

On a sequential machine, the simplest possible wrapper code is shown in Listing \ref{lst:simplest_pairloop}.
\begin{figure}
\begin{minipage}{\linewidth}
\begin{lstlisting}[language={{c}},label=lst:simplest_pairloop,caption={Pair loop in a sequential implementation}]
for (int i=0;i<npart;++i) {
  for (int j=0;j<npart;++j) {
    if (i!=j) {
     // INSERT KERNEL CODE HERE
     }
  }
}
\end{lstlisting}
\end{minipage}
\end{figure}
The computational complexity of this nested loop is $\order(N^2)$ and for short range kernels this method would be extremely inefficient. In the following we describe more advanced looping mechanisms for executing Local Particle Pair Loops on parallel architectures.
\subsection{Cell based methods for Local Particle Pair Loops}\label{sec:cell_based_methods}
If we only consider Local Particle Pair kernels with a fixed cutoff $r_c$, the computational complexity is reduced to $\order(N)$ and it is possible to use cell based looping methods (see \cite{Rapaport2004book} for an introduction).  In this approach the physical domain of size $[0,L_x]\times[0,L_y]\times[0,L_z]$ is divided into small cells of size $\Lambda_x \times \Lambda_y\times \Lambda_z$ such that $\Lambda_{x,y,z} \ge r_c$; to simplify the presentation, we assume $\Lambda=\Lambda_x=\Lambda_y=\Lambda_z$ in the following. At a given point in time every particle can be uniquely associated with one of those small cells. The local Particle Pair loop with cutoff $r_c$ can then be executed by visiting all cells $e$ in an outer loop and then iterating over all 26 neighbouring cells $e'$. Since $\Lambda\ge r_c$ it is then sufficient to consider pairs of particles $(i,j)$ such that $i\in e$ and $j\in e,e'$.

For each particle this algorithm considers potential interactions with other particles in a volume $27\Lambda^3$. However, most of these pairs will be separated by a distance $|\vec{r}^{(i)}-\vec{r}^{(j)}|>\Lambda\ge r_c$. To avoid unnecessary execution of the kernel for non-interacting particles, it is possible to add another preprocessing step which loops through all potential pairs and only stores those which are a distance of up to $\Lambda$ away. Interactions can then be calculated by looping through this neighbour list. In three dimensions this reduces the cost of the force calculation by up to a factor $81/(4\pi)\approx 6.45$. The computational overhead for building the neighbour list is usually amortised by the gain in the force calculation.

For both $\order(N)$ pair looping mechanisms described above, different \verb!ParticleDat!s can no longer be considered independently, but rather have to be seen as members of a \verb!State! object which also stores the shared cell- and neighbour lists. One particular \verb!ParticleDat! in this \verb!State! object stores the particle position, and this information is required when building the cell- or neighbour-lists. To distinguish it from other properties such as velocity and acceleration, a special derived class \verb!PositionDat! is used. As shown in Listing \ref{lst:state-example}, all \verb!ParticleDat!s in a simulation have to be associated with a \verb!State! object by setting (user-defined) properties of the state as
\begin{center}\mbox{\texttt{A.PROPERTY = ParticleDat(...)}}.\end{center} Each state also contains a domain object, which stores information about the physical domain size and boundary conditions.
\begin{figure}
\begin{minipage}{\linewidth}
\begin{lstlisting}[language={[ppmd]{python}}, label=lst:state-example,caption={Creation of a state object with position, velocity and acceleration data}]
import ppmd as md
# create state and domain objects
state = md.state.State()
state.domain = md.domain.BaseDomain()
state.domain.boundary_condition = md.domain.BoundaryTypePeriodic()
state.npart = N

# add ParticleDats to state
PositionDat = md.Data.PositionDat
ParticleDat = md.Data.ParticleDat
state.pos = PositionDat(ncomp=3,
                        dtype=c_double)
state.vel = ParticleDat(ncomp=3,
                        dtype=c_double)
state.acc = ParticleDat(ncomp=3,
                        dtype=c_double)
\end{lstlisting}
\end{minipage}
\end{figure}

During the simulation, particles will move between cells and hence if $\Lambda=r_c$ the cell- and neighbour lists need to be rebuilt at every iteration, which can be very expensive. This can be avoided by increasing the cell size and choosing an extended cutoff $\overline{r}_c$: if the relevant interaction range is $r_c < \overline{r}_c < \Lambda$ and $v_{\max}$ is the maximal particle velocity, a rebuild of the cell- and neighbour lists is only necessary every $n$ time steps if
\begin{equation}
  \overline{r}_c = r_c + 2 n\cdot\delta t\cdot v_{\max} = r_c+\delta\label{eqn:extended_cutoff}
\end{equation}
where $\delta t$ is the time step size. For time integration loops we further provide the \verb!IntegratorRange! class, which allows timestepping methods to be implemented in a way that retains the simplicity and flexibility of a standard Python \verb!range! based loop without explicit cell- and neighbour list rebuilds by the user. An example is shown in Listing \ref{lst:integrator-range}. In addition to the number of integration steps, \verb!IntegratorRange! is passed the following information:
\begin{itemize}\itemsep0ex
\item the timestep size $\delta t$,
\item a \verb!ParticleDat! containing particle velocities,
\item a maximum reuse count and
\item the thickness $\delta=\overline{r}_c - r_c$ of the additional shell.
\end{itemize}

\begin{figure}
\begin{minipage}{\linewidth}
\begin{lstlisting}[language={[ppmd]{python}}, label=lst:integrator-range,caption={Example use of \texttt{IntegratorRange} called with: \texttt{Ni} number of iterations, timestep size \texttt{dt}, velocities \texttt{v}, list reuse count \texttt{Ns} and shell thickness \texttt{delta} $=\overline{r}_c-r_c$.}]
for i in IntegratorRange(Ni,dt,v, 
                         Ns,delta):
    particle_loop_1.execute()
    force_calculation.execute()
    particle_loop_2.execute()
\end{lstlisting}
\end{minipage}
\end{figure}
\subsubsection{Parallelisation}
To simulate the interactions of a very large number of particles in a reasonable time, MD codes have to be parallelised. Modern HPC installations expose parallelism on different levels and the implication of this complex hierarchy on MD implementations will be discussed in the following. 
\paragraph{Distributed memory}
The cell-based methods described above can be parallelised with a standard domain-decomposition approach. For this the global domain is split up into smaller subdomains stored on each processor. To correctly include interactions with particles stored on neighbouring subdomains, a layer of halo cells is added. Those cells hold copies of particles which are \textit{owned} by other processors. Data in halo cells needs to be updated whenever this data changes. Note, however, that this is only necessary if the values of a particular \verb!ParticleDat! are actually read in the loop, and this information is made explicit via the access descriptors passed to the pairloop. Our code generation system will therefore only launch the corresponding parallel communication calls if necessary. This guarantees the parallel correctness of the code while avoiding superfluous and expensive parallel communications. 
 Since particles can move to different processors and hence the data layout is not fixed, there are actually two parallel communication types which need to be carried out:
\begin{enumerate}
  \item Data on particles in the halo region has to be updated if it has changed and is used in a pair loop.
  \item Particles which leave the local domain need to be moved to a different processor.
\end{enumerate}
The first operation typically needs to be performed whenever dirty data is read. For example halo exchanges on particle positions are required before every force update.  The second communication type only needs to be performed every $n$ steps, since the increased cut off in Eq. (\ref{eqn:extended_cutoff}) ensures the accuracy of the calculation even if a particle leaves the cell during those steps.  In addition to rebuilding the cell list, when a particle has left the subdomain owned by a processor, the \verb!State! object will automatically move all data owned by the particle to the receiving processor.

Virtually all modern supercomputers now consist of a large collection of relatively complex compute units (CPUs, GPUs or Xeon Phis) organised into nodes. While parallelisation between nodes is achieved with the distributed memory approach described above, each node consists of a large number of compute cores which have access to the same memory. Parallelisation across those cores on a node requires a different approach which will be described in the following section. To make use of the full machine, a hybrid approach which combines both parallelisation strategies is typically used.
\paragraph{Threading and GPU parallelisation}
To reduce memory requirements, in a sequential implementation (or if the code is parallelised purely with a distributed memory approach), the cell-list is stored as a linked list and the neighbour list is realised by storing all neighbours in a long array. This prevents any further shared memory parallelisation based on threading since neither the cell-list nor the neighbour-list can be built in parallel. To avoid this problem on GPUs we use the approach in \cite{Rapaport2011} and replace the cell list by a cell-occupancy matrix $H$. For this each particle $i$ is associated with a cell $c_i$ and the particles in a cell are arranged into ``layers'', such that all particles in a cell have a different layer-index. If the layer index of particle $i$ is $\ell_i$, then $H_{c_i,\ell_i}=i$, and $H$ can be built in parallel. Based on this, a neighbour matrix $W$ can be built such that $W_{m,i}$ is the index of the $m$-th neighbour of particle $i$. An alternative approach which is described in \cite{Rapaport1988} and avoids building $W$, would be to loop over all pairs of layers and use the matrix $H$ to identify interacting particles.

We recently also extended our framework by an OpenMP backend which is described in \cite{Saunders2017a}.
\paragraph{Vectorisation}

Modern HPC CPUs contain floating point units (FPU) which are capable of executing Single Instruction Multiple Data (SIMD) instructions. By using SIMD instructions the FPU can apply the same operation to multiple data points simultaneously. For example a 256bit wide vector FPU may simultaneously apply the same operation to four 64bit doubles or eight 32bit floats. However, producing machine code that contains these SIMD instructions is a non-trivial task. One approach to produce SIMD instructions involves explicitly implementing the desired mathematical operations using ``intrinsic'' functions for a target architecture (see e.g. \cite{Pall2013}). This ensures that SIMD instructions are generated by the compiler but requires careful implementation to be technically correct and produce efficient code. Since the intrinsics are hardware specific, this approach is not portable. In our code we currently simply avoid code patterns which inhibit auto-vectorisation by the compiler. The ability to replace loop bounds by their numerical values via \texttt{Constant} objects also helps with vectorisation. As noted in Section \ref{sec:Abstraction}, we do not currently exploit symmetry in Newton's third law when computing forces between particles (although the framework would in principle support this). We find that the Intel C/C++ compiler will successfully auto-vectorise kernels without explicitly implementing gather or scatter operations provided the kernel itself does not contain a code pattern that inhibits vectorisation. The strong- and weak- scaling results reported in Section \ref{sec:comparison} were obtained with vectorised code. We have also tried to vectorise the code by blocking pair-loops as described in \cite{Mangiardi2017}, but find that for the simple examples we considered this did not give any improvement due to additional explicit memory movement. In the future we will also explore further optimisations which are necessary for more complex kernels and consider for example a portable implementation of the vectorisation approaches in \cite{Pall2013}.
\section{Structure analysis algorithms}\label{sec:analysis}
To demonstrate that the abstraction and implementation described in the previous sections can be used to implement more complex kernels and is not restricted to force calculations, we now discuss two popular algorithms for classifying the local environment of a particle.  We show how these algorithms can be expressed in terms of particle- and local particle-pair loops. Both algorithms can be used to identify the crystalline structure of the material; an overview of other common methods can be found in \cite{Stukowski2012}.
\subsection{Bond order analysis}\label{sec:BOA}
The bond order analysis (BOA) in \cite{Steinhardt1983} introduces a set of order parameters which are defined for each particle $i$ as
\begin{equation}
  Q^{(i)}_{\ell} = \sqrt{\frac{4\pi}{2\ell+1}\sum_{m=-\ell}^{+\ell}|q^{(i)}_{\ell m}|^2}\label{eqn:Qell}
\end{equation}
with $\ell=0,1,2,\dots$. The sum
\begin{equation}
q_{\ell m}^{(i)} = \frac{1}{|\mathcal{N}(i)|}\sum_{j\in \mathcal{N}(i)} Y_{\ell}^m(\hat{\vec{r}}^{(i,j)})\label{eqn:qellm}
\end{equation}
is computed by evaluating the spherical harmonics $Y_\ell^m$ in the directions\begin{equation*}
\hat{\vec{r}}^{(i,j)}=\frac{\vec{r}^{(i)}-\vec{r}^{(j)}}{|\vec{r}^{(i)}-\vec{r}^{(j)}|}
\end{equation*}
pointing from the atom $i$ to each of its neighbours $j\in\mathcal{N}(i)$. Atoms are considered to be neighbours if their distance is smaller than a predefined cutoff range $r_c$. The moments $q_{\ell m}^{(i)}$ describe the angular dependence of the charge density $\rho^{(i)}(\vec{r}-\vec{r}^{(i)})$ of the atom's neighbours in spectral space. It can then be shown that the integral of the squared averaged charge density can be written as
\begin{equation*}
  \int_\Omega |\rho^{(i)}(\vec{r})|^2 d\Omega = \sum_{\ell=0}^{\infty} \left(Q^{(i)}_\ell\right)^2.
\end{equation*}
Perfect crystal lattices have well defined values for $Q_\ell$. In particular the order parameters with $\ell=4,5,6$ are often used to estimate the degree and nature of crystalinity.  Specific values for fcc, hcp and bcc lattices are given in Tab. \ref{tab:Q4Q6} (\cite{Stukowski2012,Mickel2013}).
\begin{table}
  \begin{center}
    \begin{tabular}{lccc}
      \hline
      Lattice Structure & $Q_4$ & $Q_5$ & $Q_6$\\
      \hline\hline
      fcc & 0.191 & 0     & 0.575\\
      hcp & 0.097 & 0.252 & 0.485\\
      bcc & 0.036 & 0     & 0.511\\
      \hline
    \end{tabular}
    \caption{Values of $Q_4$, $Q_5$ and $Q_6$ for perfect lattices, see \cite{Stukowski2012} and Tab. 1 in \cite{Mickel2013}.}
    \label{tab:Q4Q6}
  \end{center}
\end{table}
In a simulation the local structure of the material can therefore be estimated  by calculating $Q^{(i)}_\ell$ and comparing to the reference values in Tab. \ref{tab:Q4Q6}. If they agree within some tolerance, the system is classified to be in the corresponding state.

The order parameters $Q^{(i)}_\ell$ can be calculated with the two loops shown in Algorithms \ref{alg:sph_I} and \ref{alg:sph_II}. The first particle pair loop (Algorithm \ref{alg:sph_I}) calculates the number of neighbours $\nu_{\text{nb}}^{(i)}=|\mathcal{N}(i)|$ and the moments
\begin{equation*}
  \tilde{q}_{\ell m}^{(i)}= \sum_{j\in \mathcal{N}(i)} Y_{\ell}^m(\hat{\vec{r}}^{(i,j)})\quad (= \nu_{\text{nb}}^{(i)}q_{\ell m}^{(i)})
\end{equation*}
for $m=-\ell,\dots,+\ell$ for each atom $i$; those quantities are stored in two \texttt{ParticleDat}s. 
The particle loop in Algorithm \ref{alg:sph_II} uses $\nu_{\text{nb}}^{(i)}$ and $\tilde{q}_{\ell m}^{(i)}$ to calculate the $Q_{\ell}^{(i)}$ according to Eq. (\ref{eqn:Qell}); the result is stored in a third \texttt{ParticleDat}. The corresponding source code can be found in the \verb!examples/structure/boa/! subdirectory of the accompanying code release \cite{ppmd_release}.

\begin{algorithm}
\caption{BOA Local Particle Pair Loop I. \newline Input: particle positions $\vec{r}^{(i)}$ [\texttt{READ}].\newline Output: moments $q^{(i)}_{\ell m}$ [\texttt{INC\_ZERO}]}
\label{alg:sph_I}
\begin{center}
\begin{algorithmic}[1]
\FORALL{pairs $(i,j)$}
\IF{$|\vec{r}^{(i)}-\vec{r}^{(j)}|<r_c$}
  \STATE{$\hat{\vec{r}}^{(i,j)}\mapsto(\vec{r}^{(i)}-\vec{r}^{(j)})/|\vec{r}^{(i)}-\vec{r}^{(j)}|$}
  \FOR{$m=-\ell,\dots,+\ell$}
    \STATE{$\tilde{q}^{(i)}_{\ell m} \mapsto \tilde{q}_{\ell m}^{(i)}+Y_\ell^m(\hat{\vec{r}}^{(i,j)})$}
  \ENDFOR
\ENDIF
\ENDFOR
\end{algorithmic}
\end{center}
\end{algorithm}

\begin{algorithm}
\caption{BOA Particle Loop II.\newline Input: moments $\tilde{q}^{(i)}_{\ell m}$ [\texttt{READ}], number of local neighbours $\nu_{\text{nb}}^{(i)}$ [\texttt{READ}]. \newline Output: $Q_{\ell}^{(i)}$ [\texttt{WRITE}]}
\label{alg:sph_II}
\begin{center}
\begin{algorithmic}[1]
\FORALL{particles $i$}
  \FOR{$m=-\ell,\dots,+\ell$}
    \STATE{$q^{(i)}_{\ell m} \mapsto \tilde{q}_{\ell m}^{(i)}/\nu_{\text{nb}}^{(i)}$}
  \ENDFOR
  \STATE{$Q^{(i)}_\ell \mapsto \sqrt{\frac{4\pi}{2\ell+1}\sum\limits_{m=-\ell}^{+\ell}|q^{(i)}_{\ell m}|^2}$}
\ENDFOR
\end{algorithmic}
\end{center}
\end{algorithm}
\subsection{Common neighbour analysis}\label{sec:CNA}
Common neighbour analysis (CNA) \cite{Honeycutt1987} is a purely topological method for classifying the local environment of each particle. All atoms within a certain cutoff distance $r_{c}$ are considered to be ``bonded''. For any bonded pair $(i,j)$ the set of all other atoms which are bonded to both $i$ and $j$ are referred to as \textit{common neighbours}. The bonds between those common neighbours define a graph $\mathcal{G}$. For each pair $(i,j)\in\mathcal{G}$ this graph is now classified by three numbers \cite{Stukowski2012}: (1) the number of common neighbours $\nnb$, i.e. the number of vertices in $\mathcal{G}$, (2) the number of bonds $\nb$, i.e. the number of edges in $\mathcal{G}$, and (3) $\nlcb$, the number of bonds in the largest cluster (connected subgraph) $\mathcal{G}'\subset \mathcal{G}$. For each pair of bonded atoms this defines a triplet  $(\nnb,\nb,\nlcb)$ (see Fig. \ref{fig:CNA}). To classify the local environment of an atom, the triplets $(\nnb,\nb,\nlcb)$ are computed for all its neighbours and compared to reference signatures for periodic crystal structures. For example, in an hcp lattice, each atom has 12 bonds, six of which are classified as $(4,2,1)$ and the other six are $(4,2,2)$; see Tab. 1 in \cite{Stukowski2012}.
\begin{figure}
 \begin{center}
  \includegraphics[width=0.6\linewidth]{\figdir/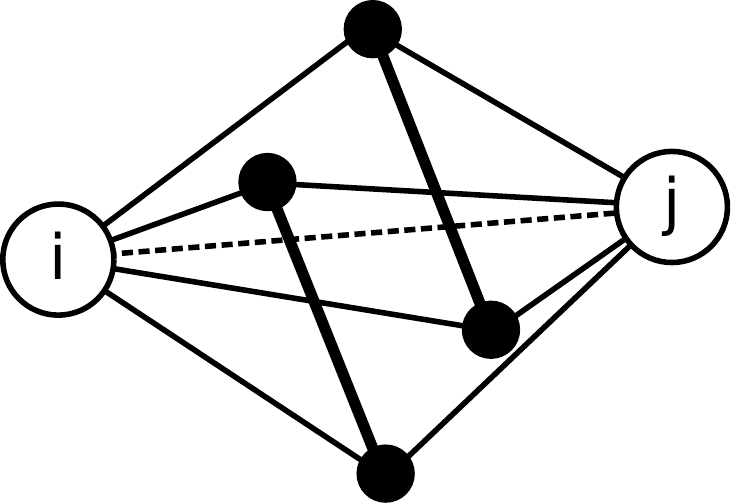}
  \caption{Common neighbour analysis for bonded atom pair $(i,j)$ (empty circles). The set of common neighbours (filled circles) are classified as a $(4,2,1)$ triplet.}
  \label{fig:CNA}
 \end{center}
\end{figure}
There is some ambiguity in the cutoff distance $r_c$. To overcome this limitation, the author of \cite{Stukowski2012} suggests an adaptive extension of the method. While this improved algorithm can also be implemented in our framework, for the sake of brevity we do not discuss this extension here and focus on the original method.

To implement the CNA algorithm in our framework we proceed in two steps: For each atom $i$ we first calculate all directly and indirectly bonded atoms. The set $\mathcal{E}^{(i)}_d$ describes the direct bonds; the indirect bonds in the local environment are collected in $\overline{\mathcal{E}}^{(i)}$ (see Fig. \ref{fig:bonds}):
\begin{equation}
 \begin{aligned}
  \mathcal{E}_d^{(i)} &= \big\{(i,v) : v \in \mathcal{N}, |\vec{r}^{(i)}-\vec{r}^{(v)}|<r_c\}\\
  \overline{\mathcal{E}}^{(i)} &= \big\{(v,w) : v,w \in \mathcal{N}, |\vec{r}^{(v)}-\vec{r}^{(w)}| < r_c,\\
  &\qquad|\vec{r}^{(i)}-\vec{r}^{(v)}| < r_c
\big\}
 \end{aligned}
\label{eqn:Ei}
\end{equation}
Since some of the indirect bonds are counted twice in $\overline{\mathcal{E}}^{(i)}$, the set 
$\mathcal{E}^{(i)}$ is an ordered representation of the same bonds:
\begin{equation}
  \mathcal{E}^{(i)} = \big\{(v,w) : (v,w) \in \overline{\mathcal{E}}^{(i)},v<w \big\}\subset \overline{\mathcal{E}}^{(i)}
\label{eqn:Ei_ordered}
\end{equation}
\begin{figure}
 \begin{center}
  \includegraphics[width=0.8\linewidth]{\figdir/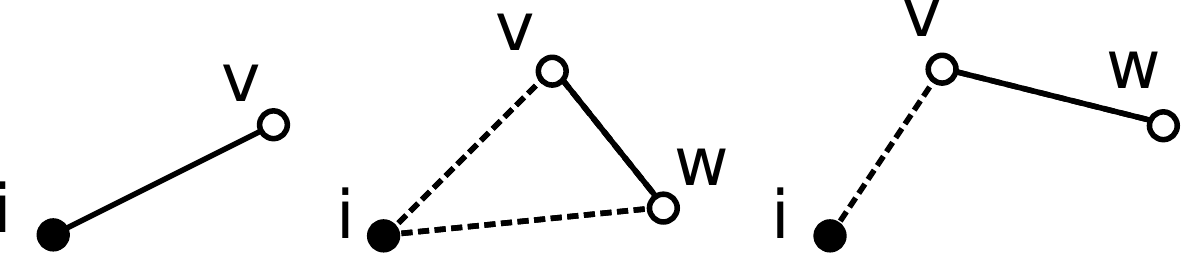}
  \caption{Example of direct (left) and indirect (centre and right) bonds as described by the sets $\mathcal{E}_d^{(i)}$, $\overline{\mathcal{E}}^{(i)}$ and $\mathcal{E}^{(i)}$ in Eqns. (\ref{eqn:Ei}) and (\ref{eqn:Ei_ordered}). The bond $(v,w)$ in the central diagram would be counted twice in $\overline{\mathcal{E}}^{(i)}$ but only once in $\mathcal{E}^{(i)}$.}\label{fig:bonds}
 \end{center}
\end{figure}\noindent
As before, $\mathcal{N}=\{0,\dots,N-1\}$ is global index set and $\mathcal{N}(i)$ the set of all neighbours of particle $i$, i.e. all other particles which are no more than a distance $r_c$ away.
In a second step we loop over all pairs $(i,j)$ of atoms and calculate the sets
\begin{equation}
 \begin{aligned}
  \mathcal{C} &= \mathcal{N}(i) \cap \mathcal{N}(j)\\
  \mathcal{E} &= \{(v,w): v,w \in \mathcal{C},v<w\} \subset \mathcal{E}^{(i)} \cap \mathcal{E}^{(j)}.
 \end{aligned}
\end{equation}
$\mathcal{C}$ is the set of common neighbours and $\mathcal{E}$ is the set of common neighbour bonds. Note that, to avoid double counting, here we consider ordered bounds $(v,w)\in\mathcal{E}^{(i)}$ such that $v<w$. Together the two sets $\mathcal{C}$ and $\mathcal{E}$ define the graph $\mathcal{G}$ introduced above. The first two entries of the triplet $(\nnb,\nb,\nlcb)$ can be calculated directly as $\nnb = |\mathcal{C}|$ and $\nb = |\mathcal{E}|$. To calculate the size of all subgraphs $\mathcal{G}'\subset\mathcal{G}$, a random node $v\in\mathcal{G}$ is chosen. The size of the subgraph $\mathcal{G}'$ such that $v\in\mathcal{G}'$ is obtained with a breadth-first traversal of the connected component containing $v$, removing all visited nodes from $\mathcal{G}$ in the process. This is repeated until all nodes have been removed, thus calculating the size of all subgraphs $\mathcal{G}'\subset\mathcal{G}$. The computation of the maximal cluster size $\nlcb=\max_{\mathcal{G}'\subset \mathcal{G}}\{|\mathcal{G}'|\}$ with this method is shown explicitly in Algorithm \ref{alg:max_cluster_size} in \ref{sec:subcluster_algorithm}.

We now show how the CNA algorithm can be implemented as a set of Local Particle Pair- loops. For this, define the following \texttt{ParticleDat}s:
\begin{itemize}
  \item $\vec{r}$ (\texttt{ncomp=3}): Particle coordinates, $\vec{r}^{(i)}$ stores the position of particle $i$
  \item $G$ (\texttt{ncomp=1}): Global id, $G^{(i)}=i\in \mathcal{N}$ stores the unique global index of particle $i$.
  \item $\nuNB$ (\texttt{ncomp=1}): Number of neighbours, i.e. $\nuNB^{(i)}=|\mathcal{N}(i)|$; this is the number of red particles in the inner circle in Fig. \ref{fig:CNA_bonds}.
  \item $\nuB$ (\texttt{ncomp=1}): Number of bonds in the local environment. $\nuB^{(i)}=|\mathcal{E}_d^{(i)}\cup\overline{\mathcal{E}}^{(i)}|$ counts the directly bonded neighbours of a particle plus the number of indirect bonds defined in Eq. (\ref{eqn:Ei}).
  \item $E$ (\texttt{ncomp=$2\nuB^{(\max)}$}): Array representation of the set $\mathcal{E}_d^{(i)}\cup\overline{\mathcal{E}}^{(i)}$ defined in Eq. (\ref{eqn:Ei}).
Two consecutive entries $E^{(i)}_{2k},E^{(i)}_{2k+1}$ represent a bonded pair in the local environment of particle $i$, i.e.\ one of the links shown in Fig. \ref{fig:CNA_bonds}. The entries of $E^{(i)}$ are arranged as follows:
\begin{itemize}
  \item $(E^{(i)}_{2k}, E^{(i)}_{2k+1}) = (G^{(i)},G^{(j)})$ with $j\ne i$ for $0\le k<\nuNB^{(i)}$
  \item $(E^{(i)}_{2k}, E^{(i)}_{2k+1}) = (G^{(j')},G^{(j'')})$ with $j'\ne i$, $j''\ne i$ for $\nuNB^{(i)}\le k< \nuB^{(i)}$
\end{itemize}
In other words, the first $\nuNB^{(i)}$ tuples represent the bonds in $\mathcal{E}_d^{(i)}$ and are shown as red (solid) lines in Fig. \ref{fig:CNA_bonds}. The remaining $\nuB-\nuNB$ tuples describe the set $\overline{\mathcal{E}}^{(i)}$ and correspond to the blue (dashed) lines. The static size $\nuB^{(\max)}$ of the list has to be chosen sufficiently large, i.e. $\nuB^{(\max)}\ge \max_i \{\nuB^{(i)}\}$.
\item $T$ (\texttt{ncomp}=$3\nuNB^{(\max)}$) stores the triplets $(\nnb,\nb,\nlcb)$ such that $(T^{(i)}_{3j},T^{(i)}_{3j+1},T^{(i)}_{3j+2})$ is the triplet $(\nnb,\nb,\nlcb)$ for the $j$-th bonded neighbour of particle $i$. The number of components $\nuNB^{(\max)}$ has to be chosen such that $\nuNB^{(\max)}\ge\max_i \{\nuNB^{(i)}\}$.
\item $t$ (\texttt{ncomp=1}) stores the number of classified bonds of particle $i$.
\end{itemize}
\begin{figure}
  \begin{center}
    \includegraphics[width=0.8\linewidth]{\figdir/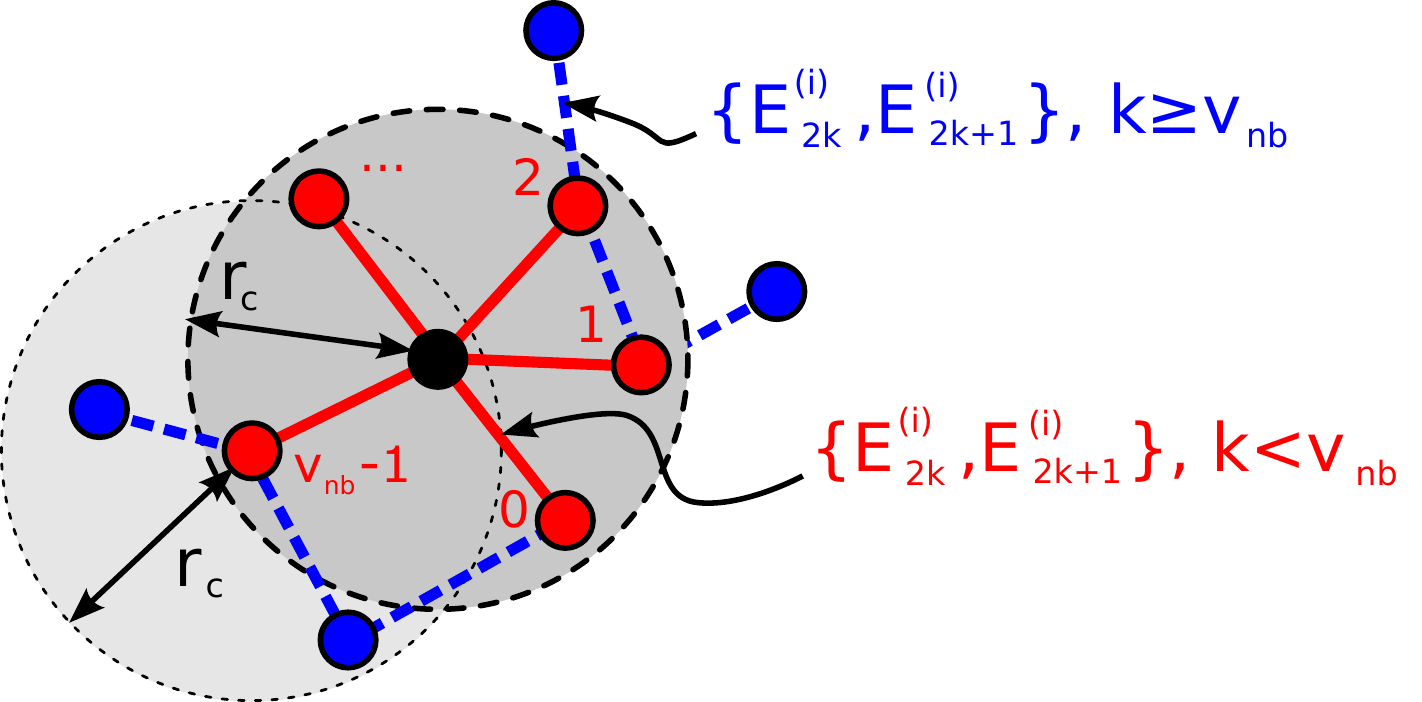}
  \end{center}
  \caption{Local bonds used for CNA construction}
  \label{fig:CNA_bonds}
\end{figure}
Using those \texttt{ParticleDat}s, for each particle the list representation $E^{(i)}$ of the set $\mathcal{E}^{(i)}_d\cup \overline{\mathcal{E}}^{(i)}$ can now be calculated with two Local Particle Pair Loops: the first loop, shown in Algorithm \ref{alg:cna_I}, calculates the first $2\nuNB^{(i)}$ entries of $E^{(i)}$ by inspecting the direct neighbours of each particle. Based on this, the second loop in algorithm \ref{alg:cna_II} adds the remaining $2(\nuB^{(i)}-\nuNB^{(i)})$ entries, i.e. the blue (dashed) lines in Fig. \ref{fig:CNA_bonds}. The final Particle Pair Loop in algorithm \ref{alg:cna_III} then uses the information stored in $E^{(i)}$ and $E^{(j)}$ to extract the tuple $(\nnb,\nb,\nlcb)$.
\begin{algorithm}
\caption{CNA Local Particle Pair Loop I: Calculate direct bonds for each particle.\newline \textit{Input}: $\vec{r}^{(i)}$ [\texttt{READ}], $G^{(i)}$ [\texttt{READ}].\newline \textit{Output}: $\nuNB^{(i)}$ [\texttt{INC\_ZERO}], $\nuB^{(i)}$ [\texttt{INC\_ZERO}],\newline$E^{(i)}$ [\texttt{WRITE}]}
\label{alg:cna_I}
\begin{center}
\begin{algorithmic}[1]
\FORALL{pairs $(i,j)$}
\IF{$|\vec{r}^{(i)}-\vec{r}^{(j)}|<r_c$}
  \STATE{$(E^{(i)}_{2\nuB},E^{(i)}_{2\nuB+1})=(G^{(i)},G^{(j)})$}
  \STATE{$\nuB^{(i)} \mapsto \nuB^{(i)}+1$}
  \STATE{$\nuNB^{(i)} \mapsto \nuNB^{(i)}+1$}
\ENDIF
\ENDFOR
\end{algorithmic}
\end{center}
\end{algorithm}

\begin{algorithm}
\caption{CNA Local Particle Pair Loop II: Calculate all other bonds in the local environment.\newline \textit{Input}: $\vec{r}^{(i)}$ [\texttt{READ}], $G^{(i)}$ [\texttt{READ}], $\nuNB^{(i)}$ [\texttt{READ}].\newline \textit{Output}: $\nuB^{(i)}$ [\texttt{INC}], $E^{(i)}$ [\texttt{RW}]}
\label{alg:cna_II}
\begin{center}
\begin{algorithmic}[1]
\FORALL{pairs $(i,j)$}
\IF{$|\vec{r}^{(i)}-\vec{r}^{(j)}|<r_c$}
  \FOR{$k=0,\dots,\nuNB^{(j)}-1$}
    \IF{$E^{(j)}_{2k+1} \ne G^{(i)}$}
      \STATE{$(E^{(i)}_{2\nuB},E^{(i)}_{2\nuB+1})=(E^{(j)}_{2k},E^{(j)}_{2k+1})$}
      \STATE{$\nuB^{(i)}\mapsto \nuB^{(i)}+1$}
    \ENDIF
  \ENDFOR
\ENDIF
\ENDFOR
\end{algorithmic}
\end{center}
\end{algorithm}

\begin{algorithm}
\caption{CNA Local Particle Pair Loop III: Calculate number of common neighbours $\nnb^{(i)}$, number of bonds $\nb^{(i)}$ between those common neighbours and the largest clustersize $\nlcb^{(i)}$.\newline \textit{Input}: $\vec{r}^{(i)}$ [\texttt{READ}], $\nuNB^{(i)}$ [\texttt{READ}], $\nuB^{(i)}$ [\texttt{READ}], $E^{(i)}$ [\texttt{READ}].\newline \textit{Output}: $T^{(i)}$ [\texttt{WRITE}], $t^{(i)}$ [\texttt{INC\_ZERO}]}
\label{alg:cna_III}
\begin{center}
\begin{algorithmic}[1]
\FORALL{pairs $(i,j)$}
\IF{$|\vec{r}^{(i)}-\vec{r}^{(j)}|<r_c$}
  \STATEx{\textit{Set $\mathcal{C}$ of common neighbours:}}
  \STATE{$\mathcal{C}\mapsto \{v:\exists k<\nuNB^{(i)}, \ell<\nuNB^{(j)},v=E^{(i)}_{2k+1}=E^{(j)}_{2\ell+1}\}$}
  \STATEx{\textit{Construct set $\mathcal{E}$ of common neighbour bonds:}}
  \STATE{$\mathcal{E}\mapsto \{\}$}
  \FOR{$k=\nuNB^{(i)},\dots,\nuB^{(i)}-1$}
    \IF{$E^{(i)}_{2k}\in \mathcal{C}$ and $E^{(i)}_{2k+1}\in \mathcal{C}$}
      \STATE{$(v,w)=(E^{(i)}_{2k},E^{(i)}_{2k+1})$}
      \IF{$w>v$} \STATE{swap $v\leftrightarrow w$} \ENDIF
      \IF{$(v,w)\ne\mathcal{E}$}
      \STATE{$\mathcal{E}\mapsto \mathcal{E}\cup(v,w)$}
      \ENDIF
    \ENDIF
  \ENDFOR
  \STATE{$T^{(i)}_{3t^{(i)}}\mapsto |\mathcal{C}|$}
  \STATE{$T^{(i)}_{3t^{(i)}+1}\mapsto |\mathcal{E}|$}
  \STATEx{\textit{Calculate largest cluster size, see Algorithm \ref{alg:max_cluster_size}:}}
  \STATE{$T^{(i)}_{3t^{(i)}+2}\mapsto\texttt{maxClustersize}(\mathcal{E})$}
  \STATE{$t^{(i)}\mapsto t^{(i)}+1$}
\ENDIF
\ENDFOR
\end{algorithmic}
\end{center}
\end{algorithm}
The C-code for Algorithms \ref{alg:cna_I} and \ref{alg:cna_II} is shown in \ref{sec:CNAkernels}. All source code (include the one for the slightly longer Algorithm \ref{alg:cna_III}) can be found in the subdirectory \verb!examples/structure/cna! of the accompanying code release \cite{ppmd_release}. Results obtained with our implementation of both a bond order- and common-neighbour-analysis algorithm are shown below in Section \ref{sec:ResultsAnalysis}.
\section{Results}\label{sec:Results}
To demonstrate the performance, portability and scalability of our code generation framework on two different chip architectures, we implemented the Velocity Verlet integrator \cite{Verlet1967} (see also e.g. \cite{Allen1989,Frenkel2001}) shown in Algorithm \ref{alg:VelocityVerlet}. Access descriptors for all loops are given in Tab. \ref{tab:access_descriptors}. The main time stepping loop is realised with an \texttt{IntegratorRange} iterator (see Section \ref{sec:cell_based_methods}), which takes care of cell-list and neighbour-list updates. C-kernels for the particle-loops that update velocity and position in lines 6 and 8 are shown in Listings \ref{lst:position_update} and \ref{lst:velocity_update}.
\begin{figure}
\begin{minipage}{\linewidth}
\begin{lstlisting}[language={{c}}, label=lst:position_update,caption={Velocity and position update kernel in the Velocity Verlet Algorithm \ref{alg:VelocityVerlet} (line 6). The constants \texttt{dt} and \texttt{dht\_iMass} are set to $\delta t$ and $\delta t/(2m)$ and passed to the pairloop as \texttt{Constant} objects.}]
v.i[0] += F.i[0]*dht_iMASS;
v.i[1] += F.i[1]*dht_iMASS;
v.i[2] += F.i[2]*dht_iMASS;
r.i[0] += dt*v.i[0];
r.i[1] += dt*v.i[1];
r.i[2] += dt*v.i[2];
\end{lstlisting}
\end{minipage}
\end{figure}
\begin{figure}
\begin{minipage}{\linewidth}
\begin{lstlisting}[language={{c}}, label=lst:velocity_update,caption={Velocity update kernel in the Velocity Verlet Algorithm \ref{alg:VelocityVerlet} (line 8). As in Listing \ref{lst:position_update}, the quantity $\delta t/(2m)$ is passed to the pairloop as a \texttt{Constant} object.}]
v.i[0] += F.i[0]*dht_iMASS;
v.i[1] += F.i[1]*dht_iMASS;
v.i[2] += F.i[2]*dht_iMASS;
\end{lstlisting}
\end{minipage}
\end{figure}
We simulated a Lennard-Jones liquid system of non-bonded particles interacting via the potential
\begin{equation}
    V(r) = 4\epsilon\left(
      \left(\frac{\sigma}{r}\right)^{12}
      -\left(\frac{\sigma}{r}\right)^{6}
      +\frac{1}{4}
    \right)\label{eqn:LJpotential}
\end{equation}
with a specified cutoff $r_c$. The C-kernel for the calculation of the resulting short-range force in line 7 is given in \ref{sec:ForceCalculation}. The full source code can be found in the \verb!code/examples/lennard-jones! subdirectory of \cite{ppmd_release}. It should be stressed that exactly the same code can be used to run the simulation both on a CPU and a GPU if the appropriate definitions shown in listing \ref{lst:GPU-switch} are added at the beginning of the Python code.
\begin{algorithm}
\caption{Velocity Verlet integrator used in Section \ref{sec:Results}. The system is integrated numerically with a time step of size $\delta t$ until the final time $T=n_{\max}\delta t$.}
\label{alg:VelocityVerlet}
\begin{center}
\begin{algorithmic}[1]
\STATE{Create \texttt{ParticleDat}s for forces $\vec{F}$ and velocities $\vec{v}$.}
\STATE{Create \texttt{PositionDat} for particle positions.}
\STATE{Initialise particle positions and velocities.}
\STATE{Collect \texttt{ParticleDat}s and \texttt{PositionDat} in a \texttt{State} object}
\FOR{timestep $i=1,\dots,n_{\max}$}
  \STATE{For all particles $i$: $\vec{v}^{(i)}\mapsto \vec{v}^{(i)} + \frac{\delta t}{2m}\vec{F}^{(i)}$, $\vec{r}^{(i)}\mapsto \vec{r}^{(i)} + \delta t\vec{v}^{(i)}$}
  \STATE{For all pairs $(i,j)$: $\vec{F}^{(i)} \mapsto \vec{F}^{(i)} + \vec{f}(\vec{r}^{(i)},\vec{r}^{(j)})$}
  \STATE{For all particles $i$: $\vec{v}^{(i)}\mapsto \vec{v}^{(i)} + \frac{\delta t}{2m}\vec{F}^{(i)}$}
\ENDFOR
\end{algorithmic}
\end{center}
\end{algorithm}
\begin{table}
\begin{center}
\begin{tabular}{lll}
\hline
Line & Loop type & Access Descriptor\\
\hline
\hline
\multirow{2}{*}{6} & \multirow{2}{*}{\texttt{ParticleLoop}} & $\vec{v}$ [\texttt{INC}], $\vec{r}$ [\texttt{INC}],\\
& &  $\vec{F}$ [\texttt{READ}], $m$ [\texttt{READ}]\\[1ex]
\multirow{2}{*}{7} & \multirow{2}{*}{\texttt{ParticlePairLoop}} & $\vec{F}$ [\texttt{INC\_ZERO}],\\
& & $\vec{r}$ [\texttt{READ}]\\[1ex]
\multirow{2}{*}{8} & \multirow{2}{*}{\texttt{ParticleLoop}} & $\vec{v}$ [\texttt{INC}], $\vec{F}$ [\texttt{READ}],\\
& &$\vec{m}$ [\texttt{READ}]\\
\hline
\end{tabular}
\caption{Access descriptors for the loops in the Velocity Verlet Algorithm \ref{alg:VelocityVerlet}.}
\label{tab:access_descriptors}
\end{center}
\end{table}
\subsection{Comparison to other codes}\label{sec:comparison}
To verify that the code generation approach does not introduce any sizable computational overheads, we compare the performance of our code to monolithic C/Fortran implementations in well established and optimised MD libraries. For this we performed the same strong scaling experiment with DL-POLY (version 4.08), LAMMPS (release dated 1$^{\text{st}}$ March 2016) and our code generation framework (subdirectory \verb!release! of \cite{ppmd_release}). Raw results can be found in the accompanying data repository \cite{paper_data}.

All codes were built with the Intel 2016 compiler suite and OpenMPI 1.8.4 (with the exception of DL-POLY, which used OpenMPI 2.0.0). The NVIDIA CUDA toolkit version 7.5.18 was used for the GPU compilation and the framework was run with Python 2.7.8. The numerical experiments were carried out on the University of Bath HPC facility ``Balena''. All nodes of the cluster consist of two Intel Xeon E5-2650v2 (2.6GHz) processors with eight cores each; in addition some nodes are equipped with Nvidia Tesla K20X GPU accelerator cards.  As the GPU port of LAMMPS offloads the force calculation, we allowed LAMMPS to use all 16 cores of the host CPU along with the GPU. In contrast, in our framework the entire simulation is run on the GPU and it is sufficient to use a single MPI rank which acts as the host controller.

We use the parameters in Tab. \ref{7.2T1}, adapted from a LAMMPS benchmark \cite{lammps_bench}. All three codes implement the neighbour list method for force calculations. For LAMMPS and our framework the extended cutoff $\overline{r}_c$ in Eq. (\ref{eqn:extended_cutoff}) was chosen such that be $\delta = \overline{r}_c - r_c = 0.1r_c$ with a neighbour list update every 20 iterations. In contrast, DL-POLY automatically updates the neighbour-list when necessary. 
\begin{table}
\centering
\begin{tabular}{ll}
\hline
Parameter & Value\\
\hline
\hline
Number of atoms: $N$             &  $10^6$ \\
Number of time steps: $n_{\max}$           & $10^4$ \\
Number density: $\rho$ & 0.8442 \\
Force cutoff: $r_c$ & 2.5 \\
Force extended cutoff: $\overline{r}_c = r_c + \delta$ & $2.75$\\
Steps between neighbour list update: & $20^\dagger$\\
\hline
\end{tabular}
\caption{Parameters of Lennard-Jones benchmark for the strong scaling experiment; units are chosen such that $\sigma=\epsilon=1$ ($\dagger = $ excluding DL-POLY, see main text).}
\label{7.2T1}
\end{table}
\begin{table*}
	\centering
	\begin{tabular}{c c c c c c}
		\hline
		Node/GPU & \multicolumn{5}{ c }{Integration Time (Seconds)} \\
		\cline{2-6}
		count & \multicolumn{2}{ c }{Framework} & \multicolumn{2}{ c }{LAMMPS} & DL\_POLY\_4 \\
			&    CPU   & GPU & CPU & GPU & CPU \\
		\hline
		\hline
		$1/16$	&	$6.83\cdot 10^{3}$	&		&	$8.22\cdot 10^{3}$	&		&	\\
		$4/16$	&	$1.49\cdot 10^{3}$	&		&	$1.67\cdot 10^{3}$	&		&	\\
		$8/16$	&	$9.18\cdot 10^{2}$	&		&	$1.05\cdot 10^{3}$	&		&	$4.99\cdot 10^{3}$\\
		$1$	&	$5.01\cdot 10^{2}$	&	$3.85\cdot 10^{2}$	&	$5.69\cdot 10^{2}$	&	$2.75\cdot 10^{2}$	&	$2.91\cdot 10^{3}$\\
		$2$	&	$2.50\cdot 10^{2}$	&		&	$2.79\cdot 10^{2}$	&		&	$1.47\cdot 10^{3}$\\
		$4$	&	$1.32\cdot 10^{2}$	&	$1.08\cdot 10^{2}$	&	$1.40\cdot 10^{2}$	&	$1.24\cdot 10^{2}$	&	$7.76\cdot 10^{2}$\\
		$8$	&	$7.50\cdot 10^{1}$	&	$6.95\cdot 10^{1}$	&	$7.32\cdot 10^{1}$	&	$6.08\cdot 10^{1}$	&	$4.92\cdot 10^{2}$\\
		$16$	&	$4.45\cdot 10^{1}$	&		&	$5.72\cdot 10^{1}$	&		&	\\
		$32$	&	$3.05\cdot 10^{1}$	&		&	$3.25\cdot 10^{1}$	&		&	\\
		$64$	&	$2.38\cdot 10^{1}$	&		&	$1.72\cdot 10^{1}$	&		&	\\
		\hline
	\end{tabular}
	\caption{Strong scaling experiment: time taken to propagate $N=10^6$ particles over $n_{\max}=10^4$ time steps.}
	\label{fig:strong_table}
\end{table*}
\begin{figure*}
\centering
    \begin{subfigure}[t]{.49\textwidth}
        \centering
        \includegraphics[width=1.0\linewidth]{\figdir/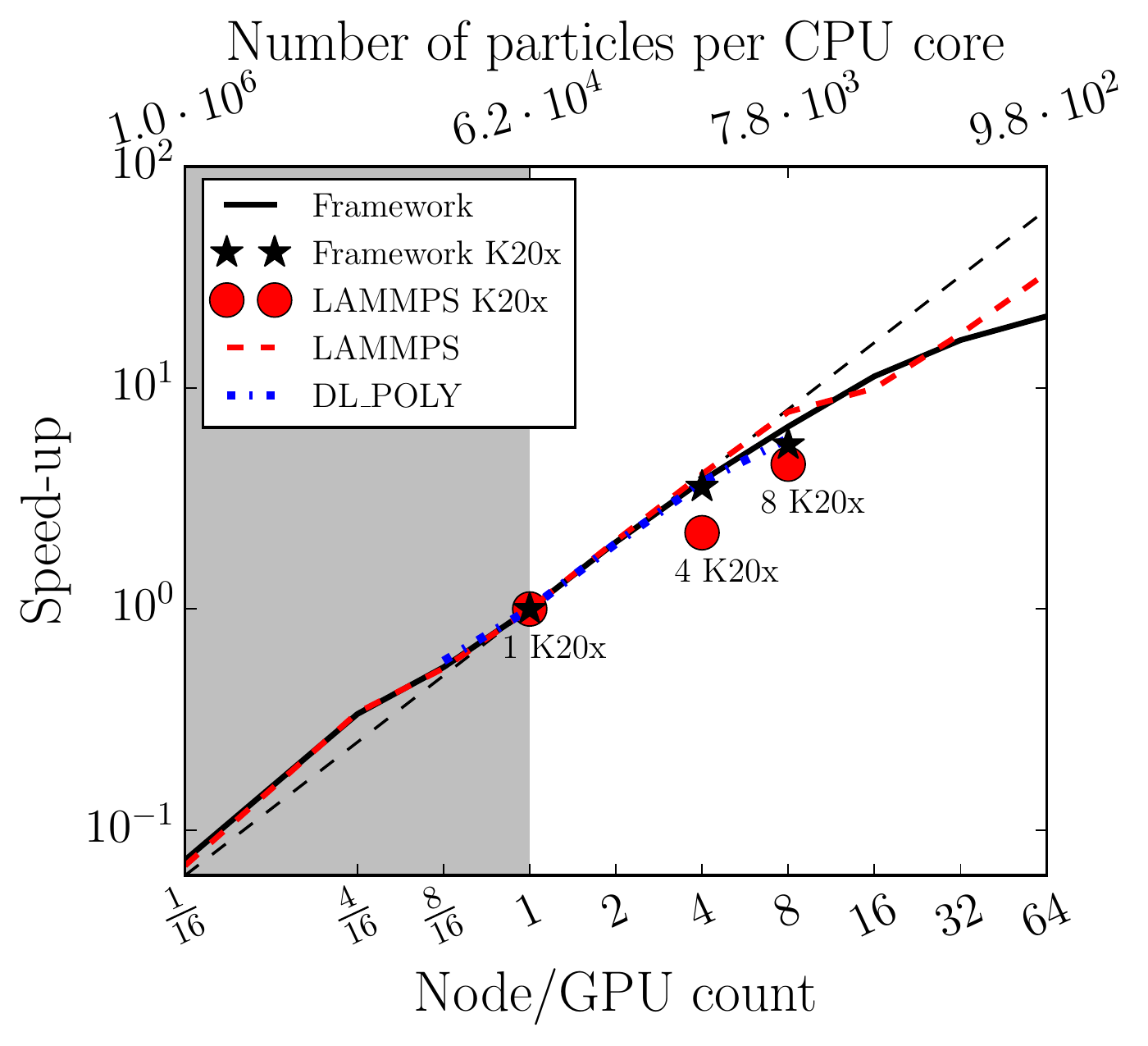}
    \end{subfigure}
    ~
    \begin{subfigure}[t]{.49\textwidth}
        \centering
        \includegraphics[width=1.0\linewidth]{\figdir/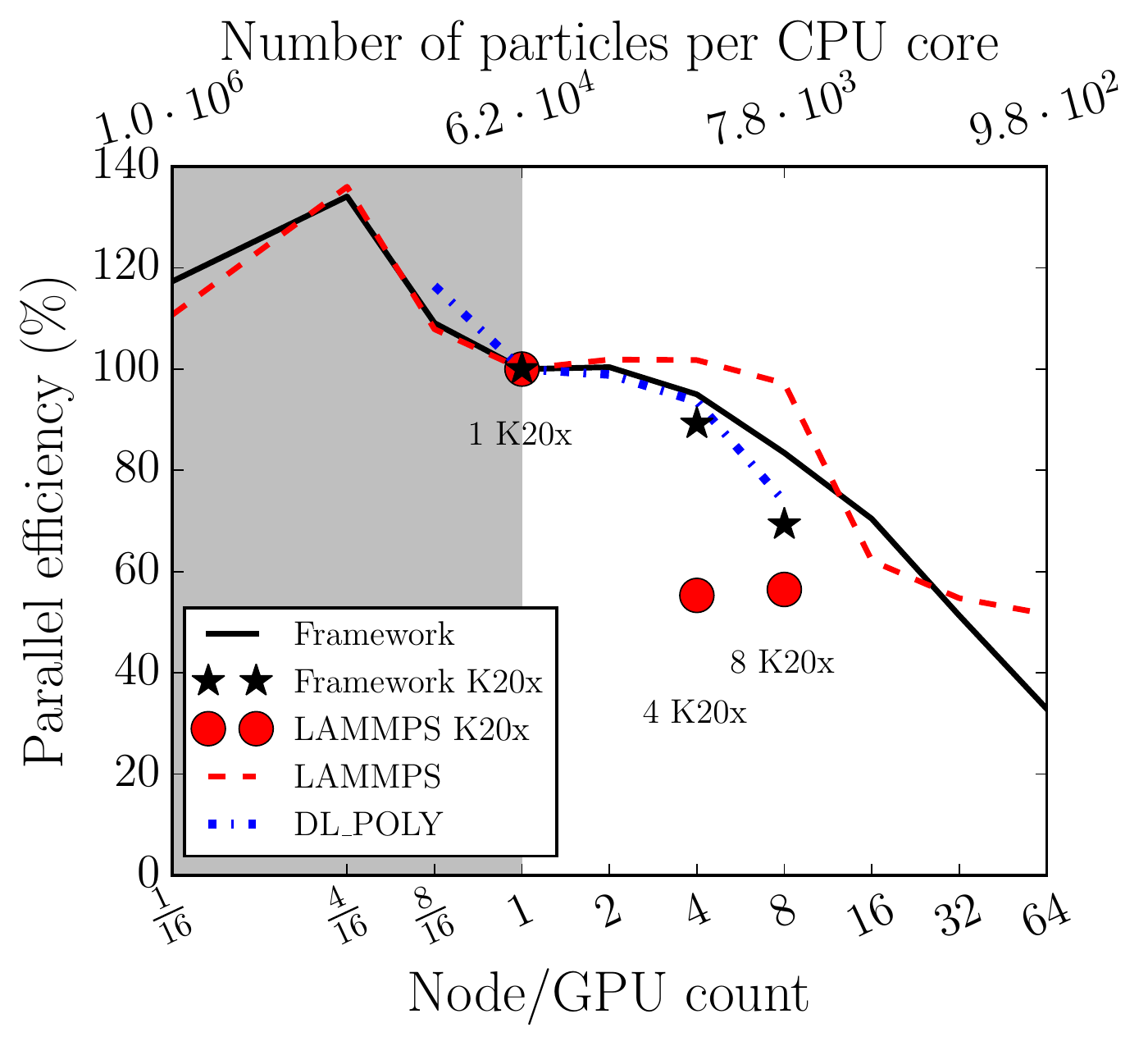}
    \end{subfigure}

    \caption{Strong scaling experiment: parallel speed-up (left) and parallel efficiency (right). Efficiency and speed-up are relative to one full node (16 cores). Efficiency is calculated according to Eqn. (\ref{eqn:strong_eff}). In the left plot perfect scaling is indicated by the dashed gray line.}
    \label{fig:strong_time}
    \label{fig:strong_eff}
\end{figure*}
The total integration time on up to 1024 cores (64 nodes) and up to 8 GPUs is tabulated in Table \ref{fig:strong_table}. Parallel speed-up and parallel efficiency are plotted in Figure \ref{fig:strong_eff}; grey regions indicate core counts contained within a single CPU node. On the largest core count (1024 cores) the average local problem size is reduced to 1,000 particles per processor. To provide a fair comparison, one K20X GPU is compared to a full 16-core CPU node since in this case the power consumption is comparable ($235\operatorname{W}$ for the K20X GPU \cite{TeslaK20xSpecs2012} vs. $2\times 95\operatorname{W}+(\text{memory power consumption})$ for the Intel Xeon E5-2650v2 CPU \cite{IntelXeonSpecs2013}). We write $t(p,N)$ for the measured wallclock time required to integrate a system with $N$ particles on $p$ CPU nodes or GPUs. The corresponding speed-up and parallel efficiency (relative to one CPU node or one GPU) are defined as
\begin{equation}
\begin{aligned}
  \text{Speed-up} &= \frac{t(1,N)}{t(p,N)}\\[1ex]
  \text{Strong parallel efficiency} &= \frac{t(1,N)}{p\times t(p,N)}
\end{aligned}
  \label{eqn:strong_eff}
\end{equation}
and shown in Fig. \ref{fig:strong_eff}.
The absolute times demonstrate that the framework provides comparable performance and scalability to DL-POLY and LAMMPS. In fact we find that for this particular setup both LAMMPS and our code are significantly faster than DL-POLY and scale better. It should be kept in mind, however, that currently both LAMMPS and DL-POLY have a much wider range of applications and provide functionality which is not yet implemented in our framework. A socket-to-socket comparison demonstrates that one full GPU can only deliver a slightly higher performance than a full CPU node. Again, the same is observed for LAMMPS. The framework can make effective use of multi-GPU systems to accelerate computation. 

To test performance for very large problem sizes we also carried out a weak scaling experiment. In this setup the average work per unit computational resource is fixed and the total problem size grows proportional to the number of nodes. A system with $512,000$ particles per CPU core ($8,192,000$ particles per node) was integrated over $5000$ timesteps. For the largest computational configuration (1024 cores) the total problem size is about half a billion ($5.24\cdot 10^8$) particles. All other system parameters are unchanged from Tab. \ref{7.2T1}. The total time for increasing problem sizes is shown in Fig. \ref{fig:weak_time} (left). The weak parallel efficiency is defined as
\begin{equation}
  \text{Weak parallel efficiency} = \frac{t(1,N)}{t(p,N\cdot p)}
  \label{eqn:weak_eff}
\end{equation}
and plotted in Fig. \ref{fig:weak_eff} (right). We observe that (relative to one node) the parallel efficiency never drops below $90\%$ and conclude that the framework will effectively scale to systems containing very large numbers of particles on a significant core count.

\begin{figure*}
	\centering
    \begin{subfigure}{.48\textwidth}
		\centering
		\begin{tabular}[b]{c c}
			\hline
			Node count & Integration Time ($10^{3}$ Seconds) \\
			\hline
			\hline
			$1/16$	&	$1.61$\\
			$2/16$	&	$1.65$\\
			$4/16$	&	$1.66$\\
			$8/16$	&	$1.52$\\
			$1$		&	$1.91$\\
			$2$		&	$1.93$\\
			$4$		&	$1.94$\\
			$8$		&	$1.96$\\
			$16$	&	$1.99$\\
			$32$	&	$2.01$\\
			$64$	&	$2.09$\\
			\hline
		\end{tabular}
    \end{subfigure}
    \begin{subfigure}{.48\textwidth}
        \includegraphics[width=1.0\linewidth]{\figdir/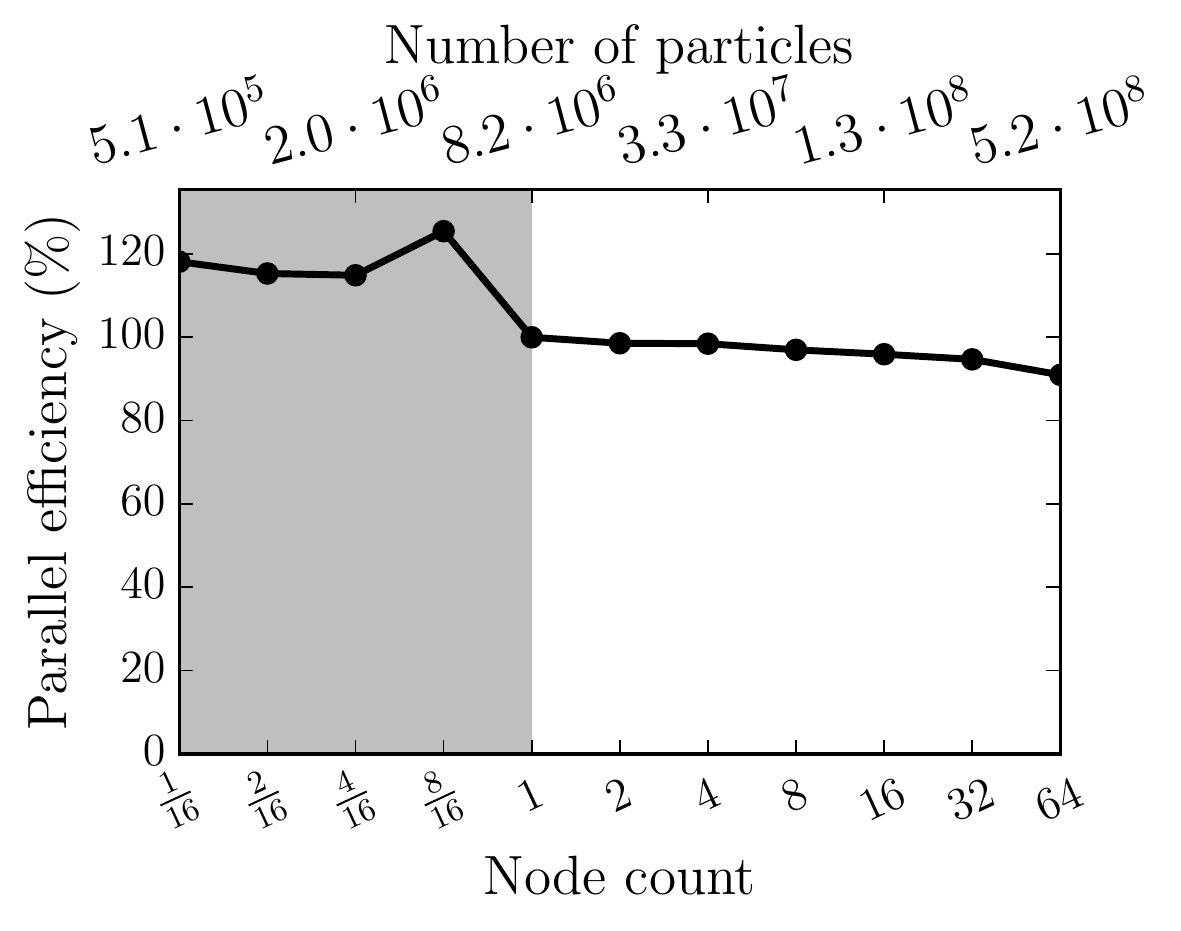}
    \end{subfigure}
    \caption{CPU-only weak scaling experiment: time taken to integrate the system over $n_{\max}=5000$ time steps (left) and parallel efficiency (right). The efficiency relative to one full node (right) is calculated according to Eqn. (\ref{eqn:weak_eff}). The top horizontal axes shows the total number $N$ of particles in the system; the number of particles per core is kept fixed at $512,000$ ($8,192,000$ particles per node).}
    \label{fig:weak_time}
    \label{fig:weak_eff}
\end{figure*}
The number of particles on a single CPU node in the previous weak scaling run is too large to fit into GPU memory. To also compare the weak scalability of the generated CPU and GPU code we therefore repeat the same experiment with a reduced number of 512,000 particles per node. The resulting time and parallel efficiency are shown in Fig. \ref{fig:gpu_comparison_time}.  While the parallel efficiency is worse for the GPU, it never drops below 60\%. On one node the GPU code is about twice as fast as the CPU code and on 16 nodes this speedup factor drops to around $1.3\times$. This can be explained by the fact that on one node the CPU implementation is slower and therefore communication overheads will have a relatively larger impact on the GPU code. To improve scalability further, we will investigate overlapping communication and communication in the future. This, however, is usually more challenging on GPUs due to the reduced work in halo regions.
\begin{figure*}
    \centering
    \begin{subfigure}{.48\textwidth}
		\centering
		\begin{tabular}[b]{c c c}
			\hline
			Node/GPU & \multicolumn{2}{ c }{Integration Time (Seconds)}\\
			\cline{2-3}
			count & CPU & GPU \\
			\hline
			\hline
				$1$	&	116.9	&	60.3\\
				$4$	&	123.2	&	78.0\\
				$8$	&	124.8	&	89.7\\
				$16$	&	129.9	&	94.1\\
			\hline
		\end{tabular}
    \end{subfigure}
    ~
    \begin{subfigure}{.48\textwidth}
        \includegraphics[width=1.0\linewidth]{\figdir/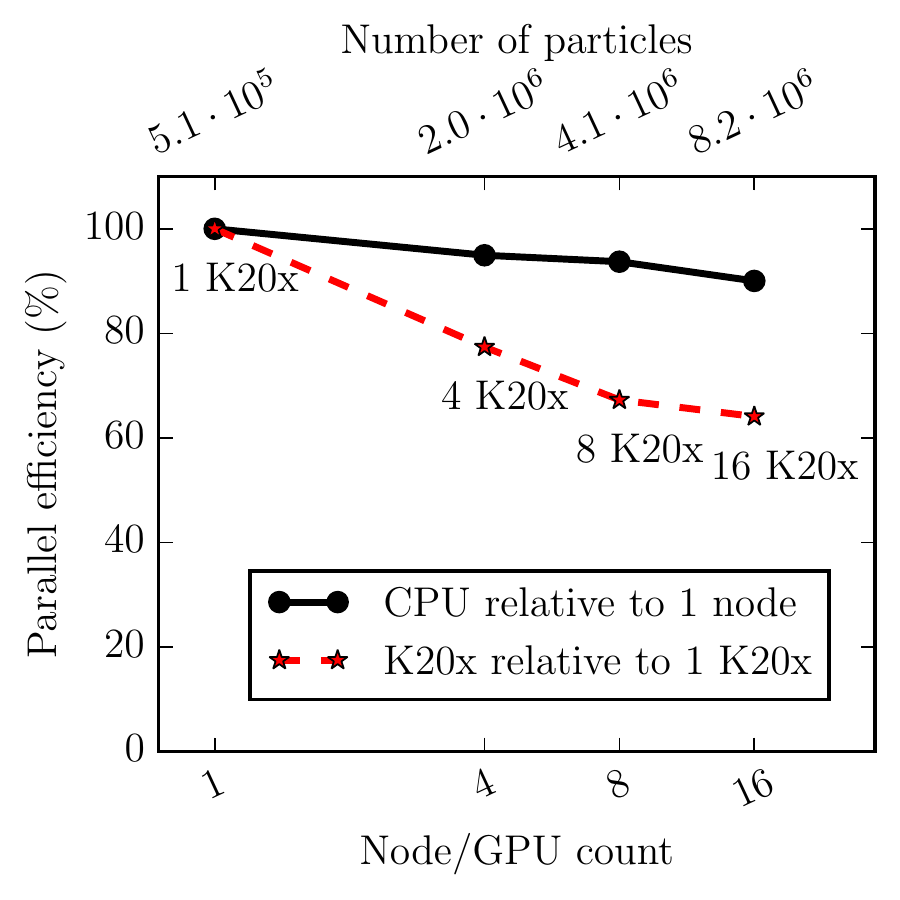}
    \end{subfigure}

    \caption{CPU-GPU weak scaling experiment with reduced particle number: time taken to simulate $n_{\max}=5000$ time steps (left) parallel efficiency relative to a single GPU/node, calculated according to Eqn. (\ref{eqn:weak_eff}) (right). The number of particles per node is kept fixed at 512,000.}
        \label{fig:gpu_comparison_eff}
        \label{fig:gpu_comparison_time}
\end{figure*}
\subsubsection{Absolute performance}
To quantify the absolute performance on both CPU and GPU we use data collected in the second weak scaling experiment (see Fig. \ref{fig:gpu_comparison_time}). The computationally most expensive operation in the simulation is the force update step performed with a particle pair loop. This accounts for $54.8\%$ of the total runtime on the CPU and $36.9\%$ on the GPU. As in this simulation the potential energy was updated every 10 iterations, we also report performance metrics for the combined force- and potential-energy (PE) update.

With the vector instruction set each core of an E5-2650v2 (2.6 GHz) Intel CPU can perform 4 double precision additions and 4 double precision multiplications per clock cycle, resulting in a total performance of 332.8 GFLOPs per node. The peak double precision floating point performance of the nVidia Tesla K20x GPU is quoted as 1.31 TFLOPs \cite{k20x}.

Absolute performance numbers for a single-node run are reported in Tab. \ref{5.2T2}. The measured times only include the time spent in the auto-generated C-code, but we found that the launch of a shared library function from Python has a negligible overhead ($\approx$ 10--20$\mu$s). Since the system is spatially homogeneous and there is little load imbalance, we report measurements collected by a single core on the fully populated node. The results demonstrate that the computationally most relevant kernels use a significant fraction of the peak floating point performance. As confirmed by the report generated by the compiler, the kernel for the Lennard-Jones force calculation in Listing \ref{lst:LJ-kernel} is automatically vectorised. 
\begin{table}
\begin{center}
\begin{tabular}{lrrrr}
  \hline
  & \multicolumn{2}{c}{Intel Xeon node} 
  & \multicolumn{2}{c}{K20X GPU}\\
  kernel & peak & time
  & peak & time\\
  \hline\hline
  Force & 16.5\% & 54.8\% & 11.9\% & 36.9\%\\
  Force \& PE & 7.5\% & 6.5\% & 14.3\% & 2.6\%\\
  \hline
\end{tabular}
\caption{Absolute performance metrics (as percentage of peak performance and integration time) for two kernels recorded from GPU weak scaling experiment presented in Fig. \ref{fig:gpu_comparison_time}. The ``Force \& PE'' kernel is only called every 10 iterations and hence accounts for a smaller proportion of the total runtime than the ``Force'' kernel.}
\label{5.2T2}
\end{center}
\end{table}
\subsection{Structure analysis algorithms}\label{sec:ResultsAnalysis}
We finally demonstrate how the structure analysis algorithms described in Section \ref{sec:analysis} can be implemented with our framework. For this we first add an on-the-fly implementation of the BOA analysis method. This is achieved by extending the main timestepping loop in Algorithm \ref{alg:VelocityVerlet} by calls to the \texttt{PairLoop} and \texttt{ParticleLoop} which evaluate $Q_\ell$ according to Algorithms \ref{alg:sph_I} and \ref{alg:sph_II}. The source code is available in the \verb!examples/on-the-fly-analysis! subdirectory of \cite{ppmd_release}.

To initialise the simulation, 125000 identical particles are arranged in a periodic cubic lattice and their velocities are sampled from a normal distribution. After allowing the system to equilibrate for 50,000 steps in an microcanonical ensemble we coupled the system to an Andersen thermostat with a target temperature near zero for 500,000 iterations. The final configuration consists of two distinct regions. The first is void of particles while the second contains a crystal structure. Fig. \ref{fig:otf_q456} shows the change of $Q_4$, $Q_5$ and $Q_6$ throughout the simulation.
\begin{figure}
 \begin{center}
  \includegraphics[width=1.\linewidth]{\figdir/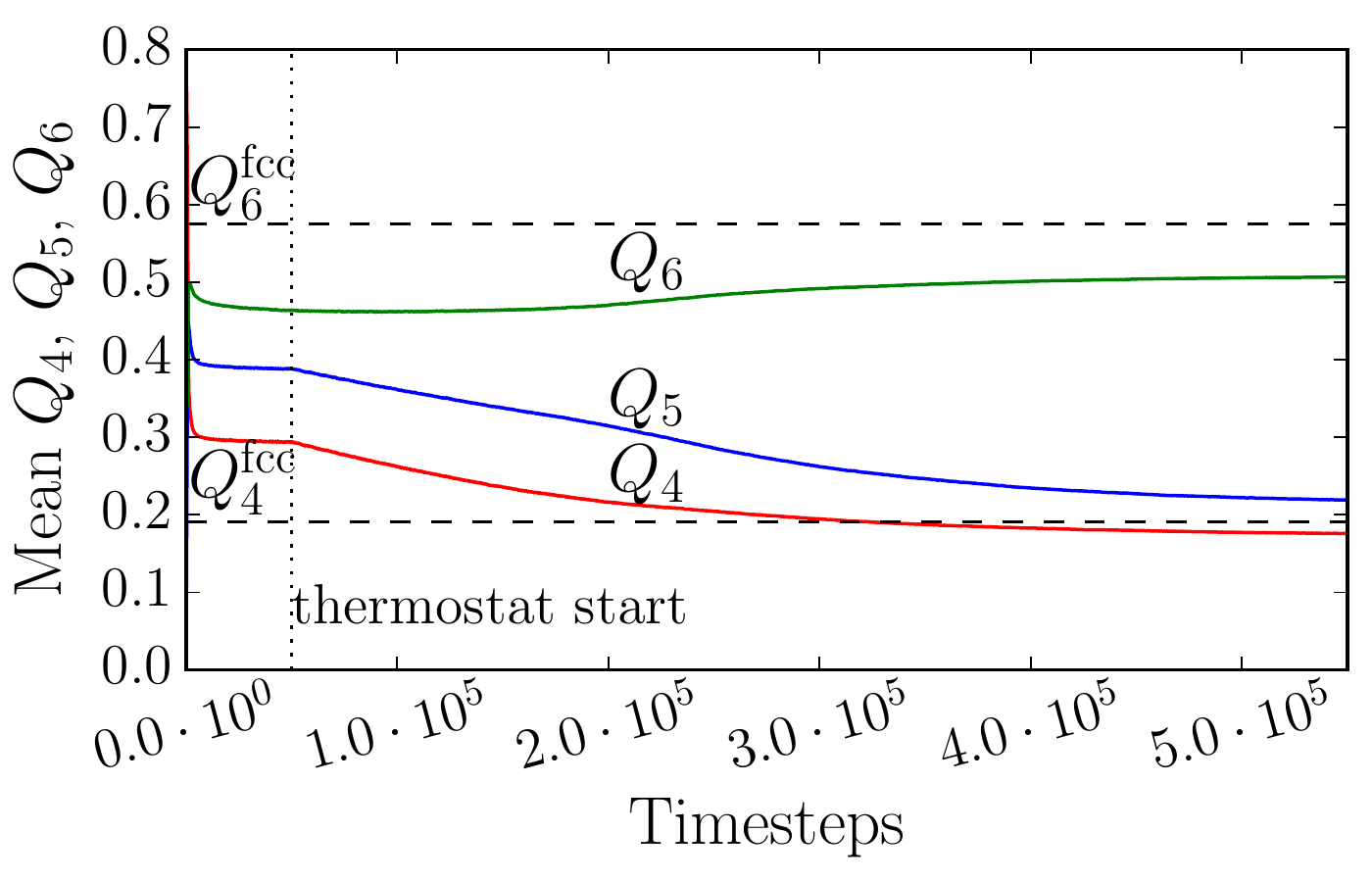}
  \caption{Evolution of mean $Q_4$, $Q_5$ and $Q_6$ values over the course of the simulation. The horizontal dashed lines plot the expected $Q_4$ and $Q_6$ values of a perfect FCC lattice.}
  \label{fig:otf_q456}
 \end{center}
\end{figure}
A distribution of the $Q_4$ and $Q_6$ values at the final timestep is shown in Figs. \ref{fig:otf_hist_q4} and \ref{fig:otf_hist_q6} in \ref{sec:appendix_BOA}. This distribution describes the proportion of FCC and HCP in the final configuration as classified by the BOA method. In this work we purely focus on the implementation of the method and do not attempt a physical interpretation of the results.

To demonstrate that the resulting code still scales well in parallel, we carry out a weak scaling experiment with the parameters in Tab. \ref{table:order_scaling}. The results are shown in Fig. \ref{fig:sph_combined} and confirm that adding the on-the-fly analysis and thermostat have no negative impact on scalability.
\begin{table}
\centering
\begin{tabular}{ll}
\hline
Parameter & Value\\
\hline
\hline
Number of atoms per node:           & $524288$ \\
Number of time steps: $n_{\max}$    & 5000 \\
Non-dimensionalised density: $\rho$ & 0.8442 \\
Force cutoff: $r_c$                 & 3.0 \\
Force extended cutoff: $\overline{r}_c = r_c + \delta$ & $3.3$\\
Steps between neighbour list updates: & $18$\\
\hline
\end{tabular}
\caption{Parameters of bond order analysis weak scaling experiment. Units are chose such that $\sigma=\epsilon=1$.}
\label{table:order_scaling}
\end{table}
\begin{figure*}
    \centering
    \begin{subfigure}{.48\textwidth}
		\centering
		\begin{tabular}[b]{c c}
			\hline
			Node count & Integration Time ($10^{2}$ Seconds) \\
			\hline
			\hline
				$1/16$	&	$4.37$\\
				$2/16$	&	$4.48$\\
				$4/16$	&	$4.50$\\
				$8/16$	&	$4.60$\\
				$1$		&	$4.99$\\
				$2$		&	$5.03$\\
				$4$		&	$5.09$\\
			\hline
		\end{tabular}
    \end{subfigure}
    \begin{subfigure}{.48\textwidth}
        \includegraphics[width=1.0\linewidth]{\figdir/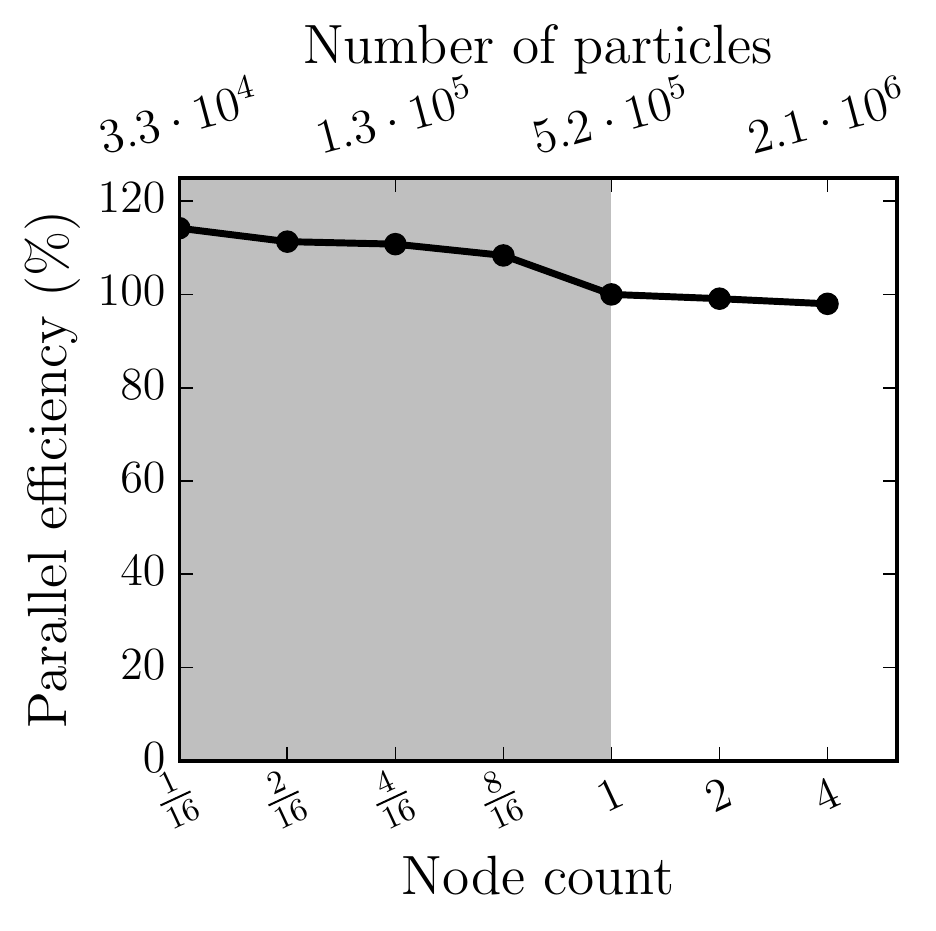}
    \end{subfigure}
    \caption{Weak scaling experiment that combines a simulation with on-the-fly analysis. Time taken to integrate 5000 steps, parallel efficiency relative to a single node (right).}
    \label{fig:sph_time}
    \label{fig:sph_eff}
    \label{fig:sph_combined}
\end{figure*}
Finally the common neighbour analysis was implemented as a parallel post-processing step. C-Kernels for Algorithms \ref{alg:cna_I}, \ref{alg:cna_II}, \ref{alg:cna_III} and \ref{alg:max_cluster_size} can be found in the subdirectory \verb!examples/structure/cna! of \cite{ppmd_release}.  We validated our implementations by verifying that perfect crystals are correctly classified in each of the FCC, BCC and HPC configurations.
We then applied the method to the test case with 125000 particles mentioned above. For the final configuration the algorithm classified 19360 (15.5\%) particles as FCC and 13052 (10.4\%) particles as HCP while 92588 (74.1\%) particles were left unclassified. Again a physical interpretation of this result would be beyond the scope of this article.
\section{Conclusions}\label{sec:Conclusion}
The key computational components of a Molecular Dynamics simulation can be expressed as loops over all particles or all particle pairs. Based on this observation, we described an abstraction for implementing those loops and introduced the necessary data structures and execution model. 
Our approach is inspired by the OP2 and PyOP2 frameworks for the solution of PDEs with grid based methods. We implemented a Python-based code generation system which allows the developer to write performance portable molecular dynamics algorithms based on a separation of concerns philosophy. By considering two popular analysis methods for the classification of crystalline structures, we showed that it is easy to apply our approach to write performant and scalable analysis code.  In principle the framework also allows for biasing dynamics within a simulation dependent on the local environment of each particle. 

The performance and scalability of our code generation framework compares favourably to two existing and well established Molecular Dynamics codes (LAMMPS and DL-POLY) both on CPUs and GPUs. This demonstrates that for the model system considered here the code generation approach does not introduce any computational overheads; the autogenerated code runs at similar speed as monolithic codes in C++ (LAMMPS) or Fortran (DL-POLY).  We stress, however, that our main aim is not to out-perform existing codes but rather explore new ways of implementing both time\-stepping methods and analysis algorithms with minimal programmer effort.

There are many ways in which our framework has to be extended to provide similar functionality to existing MD packages.
As reported in \cite{Saunders2017a}, long range force calculations with the Ewald summation method \cite{Ewald1921} are supported in a more recent version of the code; this method can be implemented directly with the data structures and looping algorithms described here. However, the computational cost of this algorithm grows with $\mathcal{O}(N^{3/2})$ and it can therefore only be used for moderate size systems. To overcome this limitation we are currently also implementing a Fast Multipole algorithm \cite{Greengard1987} which has optimal $\mathcal{O}(N)$ complexity. This approach will require new data structures such as a hierarchical mesh which stores multipole- and local- expansions in each grid cell. Since the functional form of the electrostatic interaction is fixed, long range interactions could also be simulated by linking to a standalone C-code or an existing library such as the SPME method in DL-POLY \cite{Bush2006}.
Another important extension is support for multiple species. While currently different species can be simulated by adding a species label as a \texttt{ParticleDat} and adding corresponding if-branches to the computational kernels, this is clearly not efficient and should be replaced by native support in the fundamental data structures.
Adding constraints to incorporate bonded interactions will require further work. We note, however, that excluded particles can already be treated in our framework. For this, a \texttt{ParticleDat} stores a list with global ids of all excluded particle for each atom. In the \texttt{PairLoop} kernel this exclusion list can be inspected to calculate only the relevant forces.

The performance of the GPU implementation of an algorithm is sensitive to the memory access pattern. At the beginning of a simulation particles are arranged in an ordered fashion in memory that corresponds to the physical location of the particle. As the simulation evolves the movement of particles within the simulation domain introduces an essentially random ordering of particles in memory. The results we present exhibit a slow down effect as the simulation evolves due to this sub-optimal memory ordering effect. Future versions of the framework will periodically reorder the particle data to mitigate this effect. More generally, an in-depth performance study from the perspective of memory utilisation both for the CPU and the GPU backend is important since many MD codes are memory bandwidth limited.

Finally, automatic generation of kernels from the analytical form of the potential as implemented in the OpenMM library \cite{Eastman2013} could be added. We stress, however, that it is important to still allow the user to also implement arbitrary kernels by hand to cover more general applications.
\section{Acknowledgements}
We would like to thank Alexander Stukowski (Darmstadt) for useful correspondence to clarify the exact definitions of the quantities in the common neighbour analysis algorithm. The PhD project of William Saunders is funded by an EPSRC studentship. This research made use of the Balena High Performance Computing (HPC) service at the University of Bath.
\appendix
\section{Kernels}
This appendix lists some C-kernels which are used for the Lennard-Jones force calculation in Section \ref{sec:Results} and in the common neighbour analysis discussed in Section \ref{sec:CNA}. The full source code can be found in \cite{ppmd_release}.
\subsection{Force calculation}\label{sec:ForceCalculation}
The Lennard-Jones potential in Eqn. (\ref{eqn:LJpotential}) gives rise to the force
\begin{equation}
  \begin{aligned}
   \vec{F}(\vec{r}) &= -\nabla V(r) = -\frac{\vec{r}}{r} \frac{\partial V}{\partial r} \\&= \frac{48\epsilon}{\sigma^2}\vec{r}\left(
     \left(\frac{\sigma}{r}\right)^{14}
     -\frac{1}{2}\left(\frac{\sigma}{r}\right)^8
   \right).
  \end{aligned}
  \label{eqn:LJforce}
\end{equation}
The corresponding kernel for the force- and potential calculation is shown in Listing \ref{lst:LJ-kernel} and the Python code for creating the corresponding data objects and executing the \texttt{PairLoop} is given in Listing \ref{lst:LJ-loop}. The particle position is passed in as the \verb!ParticleDat! \verb!r! and the resulting force and potential energy are returned in the \verb!ParticleDat! \verb!F! and \verb!ScalarArray! \verb!u!. The squared cutoff distance $r_c^2$ and the numerical constants $\sigma^2$, $C_V=4\epsilon$ and $C_F=-48\epsilon/\sigma^2$ are passed to the pairloop as \verb!Constant! objects. Since we use a hard cutoff, the force and potential are nonzero only if $(\vec{r}^{(i)}-\vec{r}^{(j)})^2\le r_c^2$ and only need to be calculated in this case. However, to ensure that the code can be vectorised, the force and potential is calculated for all relative distances $\vec{r}^{(i)}-\vec{r}^{(j)}$ and written to the variable \verb!F! with a ternary operator.
\begin{figure}
\begin{minipage}{\linewidth}
\begin{lstlisting}[language={{c}}, label=lst:LJ-kernel,caption={Lennard-Jones kernel}]
const double dr0 = r.i[0] - r.j[0];
const double dr1 = r.i[1] - r.j[1];
const double dr2 = r.i[2] - r.j[2];
// Calculate squared distance
// dr2 = |r_i - r_j|^2
double dr_sq = dr0*dr0+dr1*dr1+dr2*dr2;
// (sigma/dr)^2
const double r_m2 = sigma2/dr_sq;
// (sigma/dr)^4
const double r_m4 = r_m2*r_m2;
// (sigma/dr)^6 
const double r_m6 = r_m4*r_m2;
// (sigma/dr)^8
const double r_m8 = r_m4*r_m4;
// Increment potential energy
u[0]+= (dr_sq<rc_sq) ? CV*((r_m6-1.0)*r_m6+0.25) : 0.0; 
const double f_tmp=CF*(r_m6-0.5)*r_m8;
// Increment forces 
F.i[0]+= (dr_sq<rc_sq)?f_tmp*dr0:0.0;
F.i[1]+= (dr_sq<rc_sq)?f_tmp*dr1:0.0;
F.i[2]+= (dr_sq<rc_sq)?f_tmp*dr2:0.0;
\end{lstlisting}
\end{minipage}
\end{figure}

\begin{figure}
\begin{minipage}{\linewidth}
\begin{lstlisting}[language={[ppmd]{Python}}, label=lst:LJ-loop,caption={Lennard-Jones \texttt{PairLoop} implementation for the force calculation. The kernel code is defined in Listing \ref{lst:LJ-kernel}. The constants $\sigma^2$ (\texttt{sigma2}), $r_c^2$ (\texttt{rc\_sq}), $C_V=4\epsilon$ (\texttt{CV}) and $C_F=-48\epsilon/\sigma^2$ (\texttt{CF}) are passed to the kernel as \texttt{Constant} objects.}]
# Numerical constants
kernel_consts = (Constant('sigma2',
                          sigma2),
                 Constant('rc_sq',
                          rc_sq),
                 Constant('CV',
                          CV),
                 Constant('CF',
                          CF))

# Particle positions and forces
r = PositionDat(npart=npart,
                ncomp=dimension,
                dtype=c_double)
F = ParticleDat(npart=npart,
                ncomp=dimension,
                dtype=c_double)

# potential energy
u = ScalarArray(ncomp=1,
                initial_value=0.0,
                dtype=c_double)

kernel_code = ... # see Listing 9
kernel = Kernel('force',
                kernel_code,
                kernel_consts)

# Define and execute pairloop
pair_loop = PairLoop(kernel=kernel,
                     {'r':r(access.READ),
                      'F':F(access.INC),
                      'u':u(access.INC)},
                     shell_cutoff=rc)
pair_loop.execute()
\end{lstlisting}
\end{minipage}
\end{figure}
\subsection{Common Neighbour analysis}\label{sec:CNAkernels}
Computational kernels for Algorithms \ref{alg:cna_I} and \ref{alg:cna_II} in the common neighbour analysis method are shown in Listings \ref{lst:CNA-kernel_I} and \ref{lst:CNA-kernel_II}. The \verb!ParticleDat!s used in those kernels are related to the variables introduced in Section \ref{sec:CNA} and summarised in Tab. \ref{tab:cna_variables}.
\begin{table}
 \begin{center}
 \begin{tabular}{lll}
\hline
    & Description & \texttt{ParticleDat}\\
   \hline\hline
   $\vec{r}^{(i)}$ & particle position & \texttt{r}\\
   $G^{(i)}$ & global id & \texttt{id}\\
   $E^{(i)}$ & array repr. of $\mathcal{E}_d^{(i)}\cup\overline{\mathcal{E}}^{(i)}$ & \texttt{bond}\\
   $\nuNB^{(i)}$ & \# of bonded neighbours & \texttt{n\_nb}\\
   $\nuB^{(i)}$ & \# of bonds & \texttt{n\_bond}\\
\hline
 \end{tabular}
 \caption{Variables and \texttt{ParticleDat}s used in the common neighbour analysis kernels in Listings \ref{lst:CNA-kernel_I} and \ref{lst:CNA-kernel_II}.}
 \label{tab:cna_variables}
 \end{center}
\end{table}
\begin{figure}
\begin{minipage}{\linewidth}
\begin{lstlisting}[language={{c}}, label=lst:CNA-kernel_I,caption={CNA kernel for direct bond calculation in Algorithm \ref{alg:cna_I}.}]
// Calculate squared distance
const double dr0 = r.i[0] - r.j[0];
const double dr1 = r.i[1] - r.j[1];
const double dr2 = r.i[2] - r.j[2];
double dr_sq = dr0*dr0+dr1*dr1+dr2*dr2;
if (dr_sq < rc_sq) {
  // Add direct bond
  bond.i[2*n_bond.i[0]] = id.i[0];
  bond.i[2*n_bond.i[0]+1] = id.j[0];
  // Increment number of neighbours
  n_nb.i[0]++;
  // Increment number of bonds
  n_bond.i[0]++;
}
\end{lstlisting}
\end{minipage}
\end{figure}
\begin{figure}
\begin{minipage}{\linewidth}
\begin{lstlisting}[language={{c}}, label=lst:CNA-kernel_II,caption={CNA kernel for indirect bond calculation in Algorithm \ref{alg:cna_II}.}]
// Calculate squared distance
const double dr0 = r.i[0] - r.j[0];
const double dr1 = r.i[1] - r.j[1];
const double dr2 = r.i[2] - r.j[2];
double dr_sq = dr0*dr0+dr1*dr1+dr2*dr2;
if (dr_sq < rc_sq) {
  for (int k=0;k<n_nb.j[0];++k) {
    // Add indirect bond
    if (bond.j[2*k+1] != id.i[0]) {
      bond.i[2*n_bond.i[0]] = bond.j[2*k];
      bond.i[2*n_bond.i[0]+1] = bond.j[2*k+1];
      // Increment number of bonds
      n_bond.i[0]++;
    }
  }
}
\end{lstlisting}
\end{minipage}
\end{figure}
\section{Key Variables}
A list of key physical variables used in this paper can be found in Tab. \ref{tab:variables}.
\begin{table}
\begin{center}
\begin{tabular}{ll}
\hline
Variable & Definition\\
\hline
\hline
$\vec{r}$ & position\\
$\vec{v}$ & velocity\\
$v_{\max}$ & maximal velocity\\
$\vec{F}$ & force\\
$m$ & mass\\
$V$ & potential\\
$\delta t$ & time step size\\
$r_c$ & cutoff distance\\
$\overline{r}_c$ & extended cutoff (see Eq. (\ref{eqn:extended_cutoff}))\\
$N$ & number of particles\\
\hline
\end{tabular}
\caption{Key variables used in this paper.}
\label{tab:variables}
\end{center}
\end{table}
\section{Bond Order Analysis}\label{sec:appendix_BOA}
Figures \ref{fig:otf_hist_q4} and \ref{fig:otf_hist_q6} show the final distribution of the order parameters $Q_4$ and $Q_6$ in the numerical experiment described in Section \ref{sec:ResultsAnalysis}.
\begin{figure}
 \begin{center}
  \includegraphics[width=1.\linewidth]{\figdir/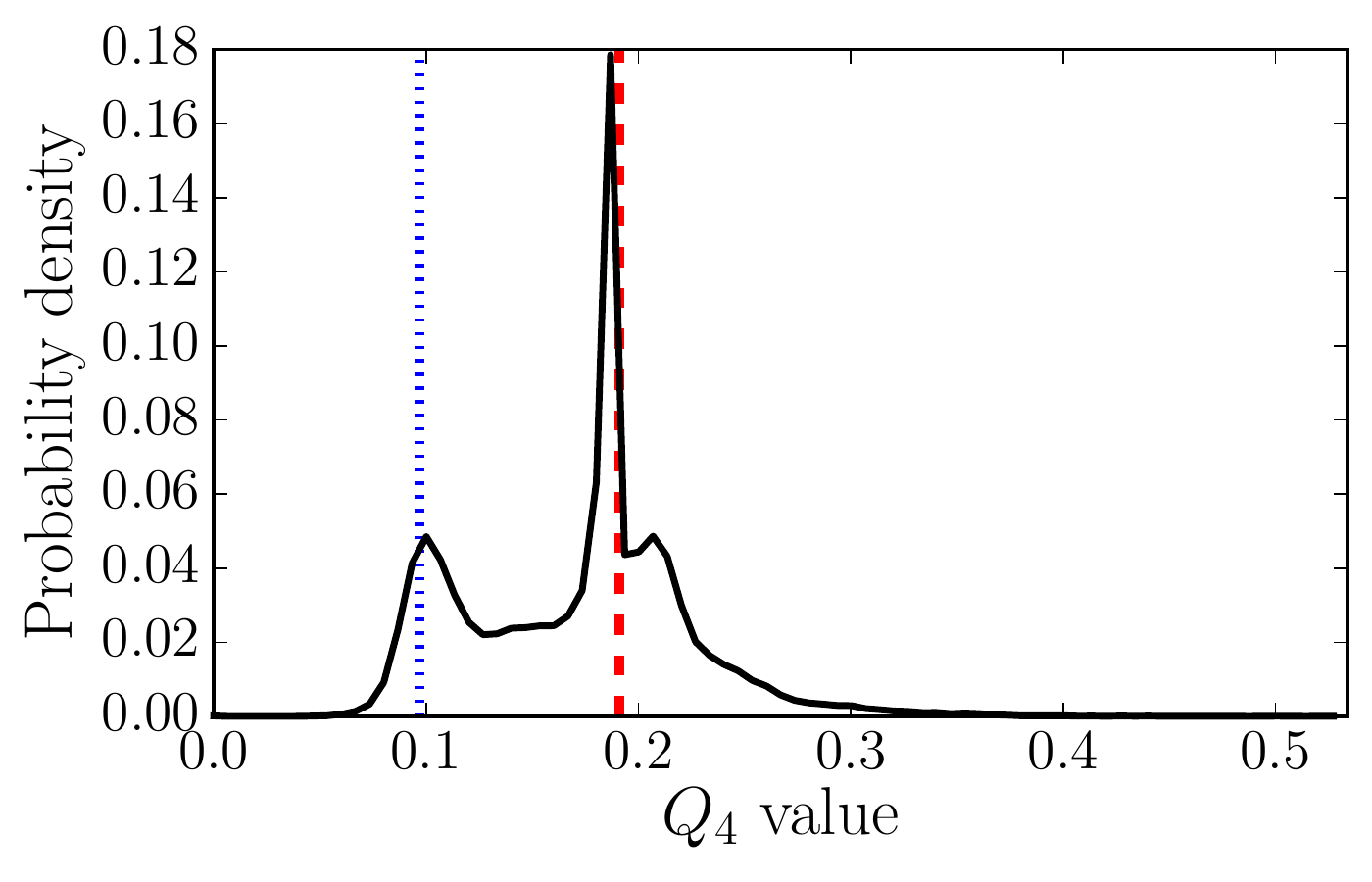}
  \caption{Probability density of $Q_4$ values in final system configuration. Dashed vertical line at $Q_4=0.097$ is the expected $Q_4$ value of a perfect hcp lattice. Dashed vertical line at $Q_4=0.191$ is the expected $Q_4$ value of a perfect fcc lattice. }
  \label{fig:otf_hist_q4}
 \end{center}
\end{figure}

\begin{figure}
 \begin{center}
  \includegraphics[width=1.\linewidth]{\figdir/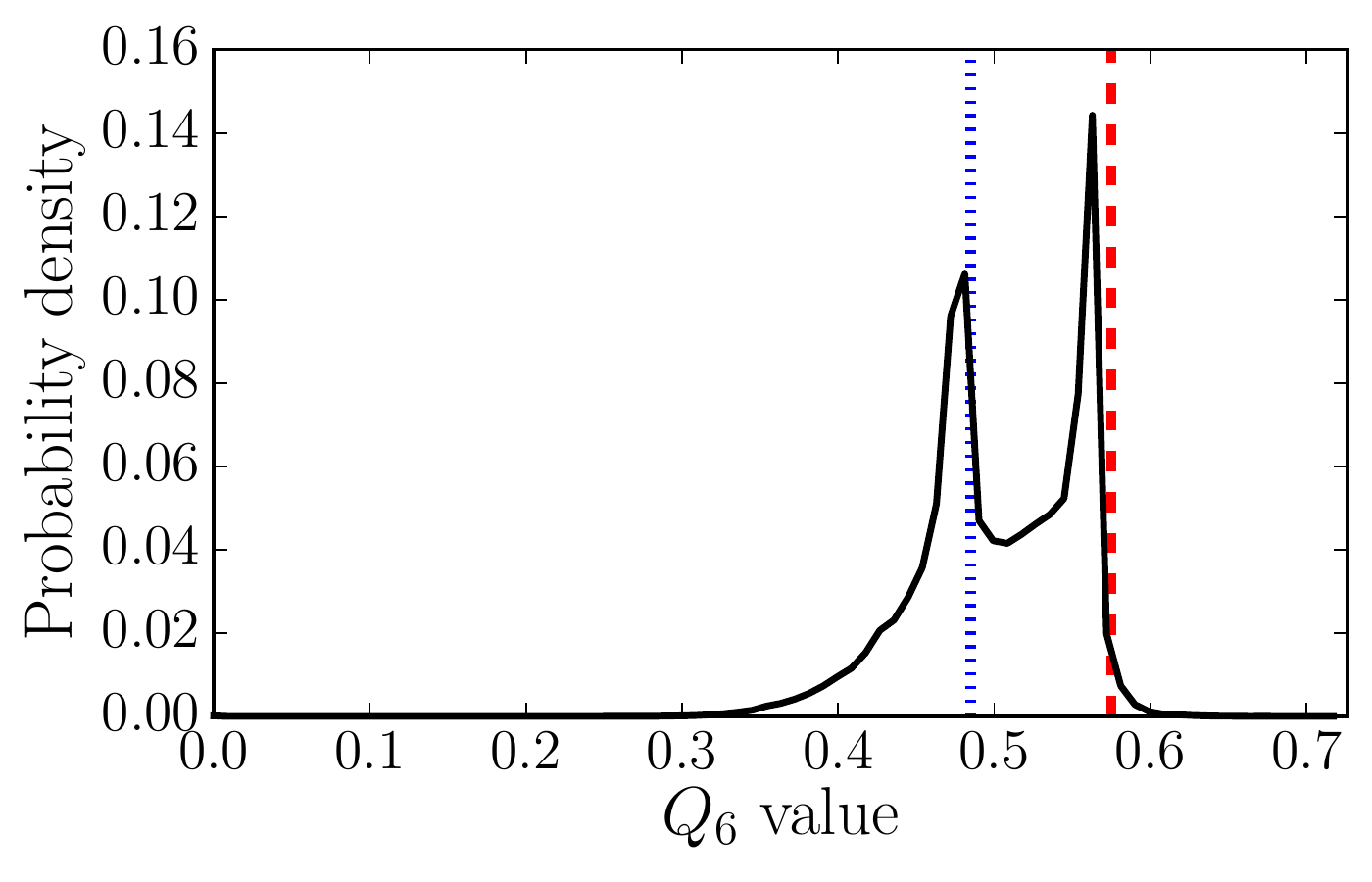}
  \caption{Probability density of $Q_6$ values in final system configuration. Dashed vertical line at $Q_6=0.485$ is the expected $Q_6$ value of a perfect hcp lattice. Dashed vertical line at $Q_6=0.575$ is the expected $Q_6$ value of a perfect fcc lattice. }  \label{fig:otf_hist_q6}
 \end{center}
\end{figure}
\section{Largest subcluster algorithm}\label{sec:subcluster_algorithm}
Algorithm \ref{alg:max_cluster_size} can be used to calculate the size of the largest connected component of a graph given by a set of edges $\mathcal{E}$. For this the edges in each subgraph are counted with a breadth-first traversal, counting and removing all visited edges in the process.
\begin{algorithm}
\caption{Calculate maximal cluster size.\newline \textit{Input}: graph defined by a set of edges $\mathcal{E}$.\newline \textit{Output}: $S_{\max}$, the size of the largest cluster}
\label{alg:max_cluster_size}
\begin{center}
\begin{algorithmic}[1]
\STATE{$S_{\max}\mapsto 0$}
\WHILE{$\mathcal{E}\ne\emptyset$}
  \STATE{$S\mapsto 0$}
  \STATE{Pick some edge $(v_1,v_2)\in \mathcal{E}$}
  \STATE{$\mathcal{Q}\mapsto \{v_1\}$}
  \WHILE{$\mathcal{Q}\ne\emptyset$}
    \STATE{Pick some $v\in \mathcal{Q}$ and remove it from $\mathcal{Q}$}
    \STATE{$\mathcal{P}\mapsto \{(v,w)\in \mathcal{E}\}$}
    \STATE{$\mathcal{Q}\mapsto \mathcal{Q} \cup \{w:(v,w)\in \mathcal{P}\}$}
    \STATE{$S\mapsto S+|\mathcal{P}|$}
    \STATE{Remove all edges $e\in \mathcal{P}$ from $\mathcal{E}$}
  \ENDWHILE
  \STATE{$S_{\max}\mapsto \max\{S,S_{\max}\}$}
\ENDWHILE
\end{algorithmic}
\end{center}
\end{algorithm}
\FloatBarrier
\ifbool{PREPRINT}{ 
}{} 
\bibliographystyle{elsarticle-num}

\end{document}